%% file: main.tex
\documentclass[useAMS,usenatbib]{mn2e}

\usepackage{graphicx}
\usepackage[]{natbib}
\usepackage[]{subfig}
\usepackage{amssymb}
\usepackage{grffile}

\newcommand\msun{ M$_\odot$}%
\newcommand\hmsun{ $h^{-1}$M$_\odot$}%
\newcommand\kpc{pkpc}%
\newcommand\smr{size - mass relation}%
\newcommand\bigbox{Ref-L100N1504}

\newcommand\highres{Recal-L025N0752}
\newcommand\highresref{Ref-L025N0752}
\newcommand\Eagle{{\sc EAGLE}}
\newcommand\Furlong{\cite{Furlong15}}
\newcommand\Schaye{\cite{Schaye15}}
\newcommand\Gadget{{\sc Gadget}}
\newcommand\Anarchy{{\sc Anarchy}}

\newcommand\Subfind{{\sc SUBFIND}}
\newcommand\skirt{{\sc skirt}}
\newcommand\OWLS{{\sc OWLS}}

\newcommand\GIMIC{{\sc GIMIC}}
\newcommand\Illustrius{{\sc ILLUSTRIS}}

\newcommand\Cloudy{{\sc Cloudy}}

\newcommand\Candels{{\sc CANDELS}}
\newcommand\sersic{S\'{e}rsic}
\newcommand\vdw{VDW14}
\newcommand\shen{\cite{Shen03}}

\newcommand\tht{$^{\rm th}$}
\newcommand\highz{high-$z$}
\newcommand\hmr{half-mass radius}
\newcommand\hmrz{$R_{50, {\rm stars}(z=2)}(z=0)$}

\title[\Eagle: Evolution of galaxy sizes]{Size evolution of normal and compact galaxies in the \Eagle\ simulation}
\author[Furlong et al]{ 
M. Furlong$^1$\thanks{E-mail:michelle.furlong@durham.ac.uk},  R. G. Bower$^1$, R. A. Crain$^2$, J. Schaye$^3$, T. Theuns$^{1}$, J. W. Trayford$^1$,
\newauthor
 Y. Qu$^1$, M. Schaller$^1$, M. Berthet$^1$ and J. C. Helly$^1$ 
\newauthor
\\
$^{1}$Institute for Computational Cosmology, Durham University, U.K. \\
$^{2}$Astrophysics Research Institute, Liverpool John Moores University, 146 Brownlow Hill, Liverpool L3 5RF  \\
$^{3}$Leiden Observatory, Leiden University, P.O. Box 9513, 2300 RA Leiden, the Netherlands \\}

\begin{document}

\date{Accepted XXXX  Received XXXX; in original form XXXX}

\pagerange{\pageref{firstpage}--\pageref{lastpage}} \pubyear{2015}

\maketitle

\label{firstpage}

\begin{abstract}
We present the evolution of galaxy sizes, from redshift 2 to 0, for actively star forming and passive galaxies in the cosmological hydrodynamical 100$^3$ cMpc$^3$ simulation of the \Eagle\ project.
We find that the sizes increase with stellar mass
, but that the relation weakens with increasing
redshift. Separating galaxies by their star formation activity, we find
that passive galaxies are typically smaller than active galaxies at fixed stellar mass. 
These trends are consistent with those found in observations and the
level of agreement between the predicted and observed \smr\ is of order 0.1 dex for
$z<1$ and 0.2--0.3 dex from redshift 1 to 2. We use the simulation to
compare the evolution of individual galaxies to that of 
the population as a whole. While the evolution of
the size-stellar mass relation for active galaxies provides a
good proxy for the evolution of individual galaxies, the evolution of individual passive galaxies is not well represented by the observed \smr\ due to the evolving number density of passive galaxies. 
Observations of $z\sim2$ galaxies have revealed an 
abundance of massive red compact galaxies, that depletes below $z\sim1$. We find that a similar population forms naturally in the simulation.
Comparing these galaxies to their $z=0$ descendants, we find that all compact galaxies grow in size due to the high-redshift stars migrating outwards.
Approximately 60\% of the compact galaxies increase in size further due to renewed star formation and/or mergers.
\end{abstract}


\input{Introduction}

\input{Simulations}

\input{Results}

\input{Conculsions}

\section*{Acknowledgements}
This work used the DiRAC Data Centric system
at Durham University, operated by the Institute for Computational Cosmology on behalf of the STFC DiRAC HPC
Facility (www.dirac.ac.uk). This equipment was funded by
BIS National E-infrastructure capital grant ST/K00042X/1,
STFC capital grant ST/H008519/1, and STFC DiRAC
Operations grant ST/K003267/1 and Durham University.
DiRAC is part of the National E-Infrastructure. We also
gratefully acknowledge PRACE for awarding us access to
the resource Curie based in France at Tr\'es Grand Centre de Calcul. 
This work was sponsored by the Dutch National Computing Facilities Foundation (NCF) for the use
of supercomputer facilities, with financial support from the
Netherlands Organization for Scientific Research (NWO).

RAC is a Royal Society University Research Fellow.
The research was supported in part by the Interuniversity Attraction Poles
Programme initiated by the Belgian Science Policy OWNce
([AP P7/08 CHARM]), the National Science Foundation under Grant No. NSF PHY11-25915, 
the UK Science and Technology Facilities Council (grant numbers ST/F001166/1 and
ST/I000976/1), Rolling and Consolidating Grants to the
ICC, Marie Curie Reintegration Grant PERG06-GA-2009-256573 and ERC grant agreement 278594-GasAroundGalaxies.

\bibliographystyle{mn2efix} 
\bibliography{sizebib}

\appendix
\input{Appendix}

\bsp

\label{lastpage}

\end{document}

%% file: Introduction.tex
\section{Introduction}

Understanding and reproducing the observed evolution of galaxies is one of the key aims of modern cosmology. 
Galaxies are thought to form as a result of gas cooling in the gravitational potential dominated by dark matter \citep[e.g.][]{White78, White91}. As the baryons cool, they form a rotating disk.
Assuming the gas initially has a specific angular momentum similar to that of the dark matter and that angular momentum is conserved, the $\Lambda$CDM framework predicts the evolution of the sizes of disk galaxies \citep{Fall80, Mo98}. 
In practice, however, the process of baryon condensation is complicated by the formation of stars and black holes and the energy that they feed back into the surrounding gas. 
The angular momentum of the stars that form 
may therefore differ significantly from that of the halo.  The evolution of galaxy sizes is therefore a critical 
test of the physical processes of galaxy formation.

Recent observational surveys of the $z\sim 0$ Universe 
sample millions of galaxies, probing the full range of galaxy masses and sizes \citep[e.g.][]{Shen03, Dutton11, Baldry12}. The evolution of the sizes 
of galaxies from the present day up to redshift 3 has been mapped for a broad range of galaxy masses and types using the Hubble Space Telescope with \Candels\ imaging \citep[ hereafter \vdw]{Patel13, vanDokkum14, vanderWel14}. 
Galaxies in the distant Universe are smaller at fixed mass than local systems, suggesting
that galaxies undergo significant size evolution. 

At $z \gtrsim 1$, observations reveal a class of very compact passive galaxies, often referred to as ``red nuggets'' \citep[e.g.][]{Cimatti04, Daddi05, Trujillo06}.  Such galaxies are rare in the nearby Universe \citep[][\vdw]{Poggianti13}. 
The evolutionary path of individual compact galaxies cannot be inferred from the observational data alone since galaxies evolve in both stellar mass and size, and may merge with other systems.
It is thus challenging to establish what becomes of these compact passive galaxies from observational evidence alone.
Simulations, such as the one considered here, can provide the necessary link to the present-day galaxy population.

Until recently, the sizes of galaxies produced by hydrodynamical simulations were not consistent with observations: while the 
masses of the galaxies tended to be too large, their sizes were too small \citep[e.g.][]{Aquilla}. The dense gas transfers angular momentum to the outer halo, which is exacerbated if too much gas is turned into stars, leading to what became known as the angular momentum catastrophe \citep{Katz&Gunn91, Navarro&White94}.
This problem can be overcome by the inclusion of energetic feedback, which tends to expel the lowest angular momentum gas from the forming galaxy, while also regulating the stellar mass formed \citep[e.g.][]{Sales10, Brook11, Brook12a}. 
Including realistic and efficient feedback has resulted in more reasonable galaxy sizes for disk galaxies in small galaxy samples or over specific mass ranges \citep{Governato04, Okamoto05, Sales10, Brook12, McCarthy12, Munshi13, Aumer13, Hopkins14, Marinacci14}.
However, reproducing the observed sizes of the general galaxy population remains a challenge for many cosmological simulations.
The production of realistic galaxy stellar masses in such simulations, or indeed a broad range of galaxy scaling relations, is a necessary, but not sufficient, criterion for the reproduction of galaxy sizes \citep{Aquilla, Crain15}. 

The distribution of galaxy sizes has been studied in cosmological simulations in the work of \cite{Sales10} and \cite{McCarthy12}, who focused on the $z=2$ and $z=0$ populations in the \OWLS\ \citep{Schaye10} and \GIMIC\ \citep{Crain09} simulations respectively.
A recent study based on the \Illustrius\ simulation \citep{Illustrius} by \cite{Snyder15} examined the size-mass-morphology relation at $z=0$.
The galaxy sizes were shown to be a factor of $\sim 2$ too large relative to observations at this redshift and the evolution of sizes was not examined.
Using the same simulation, \cite{Wellons15, Wellons15b} looked at the formation and evolution of $z=2$ massive compact galaxies respectively.
The size evolution of the compact galaxies was found to depend on the environment in which the galaxies are found.
We will also examine the evolution of compact galaxies, but in the broader context of the \highz\ passive population in the simulation, after comparing the full galaxy population to observations, finding good agreement.

In this study we investigate the evolution of the galaxy \smr\ and evolutionary tracks of individual galaxies using the Evolution and Assembly of Galaxies and their Environments (\Eagle) simulation suite \citep{Schaye15, Crain15}.
The largest of the \Eagle\ simulations evolves a volume of 100$^3$ comoving Mpc$^3$ (cMpc$^3$), with 13,200 galaxies of stellar mass $M_* > 10^9$\msun\ and 3,562 with $M_*>10^{10}$\msun\ at $z=0$.
This sample enables us to study the evolution of galaxy sizes from $z=2$ to the present day, while separating the population into those that are star forming and passive. 

The simulation combines gravitational N-body solvers with state-of-the-art smoothed particle hydrodynamics and subgrid models for
the formation of stars, metal-dependent cooling, stellar mass loss and energy injection from stars and black holes.
The simulation reproduces a wide range of properties of observed galaxies with unprecedented fidelity \citep{Schaye15}. 
The observed {\it present-day} galaxy stellar mass function, galaxy disk sizes and black hole masses were used to calibrate the uncertain parameters of the subgrid components, but other aspects of the model, such as the star formation rates \citep{Schaye15}, galaxy colours \citep{Trayford15}, rotation curves \citep{Schaller15a}, the gas content of galaxies \citep{Lagos15, Bahe15}, and the inter-galactic hydrogen and metal absorption lines \citep{Schaye15, Rahmati15}, as well as the evolution of the galaxy stellar mass function and star formation rates \citep{Furlong15} represent predictions of the simulation and exhibit broad agreement with observations.  
A full discussion of the model calibration/validation philosophy can be found in \Schaye\ and \cite{Crain15}.  

Although the $z=0$ galaxy disk sizes were considered during the calibration of the \Eagle\ simulation, the {\it evolution} of galaxy sizes is a prediction of the simulation. Moreover, the simulations allow us to explore the difference in size between galaxies that are actively forming stars and those that are passive, a distinction that was not made during the calibration. Passive galaxies are particularly prevalent at high stellar masses, and primarily grow in mass through galaxy mergers. 
Such galaxies are observed to be smaller than disk galaxies of the same stellar mass \citep[e.g.][]{Shen03, Baldry12, vanderWel14} providing a key test of the simulation.
This comparison has further implications than simply validating the model since the simulation can also help us interpret the observational data. For example, we can trace galaxies as they evolve, enabling us to address issues such as the origin of the \smr\ and to relate galaxy masses and sizes at one redshift to another redshift.  
In particular, we will show that the simulation produces a population of high-redshift compact galaxies, and we trace the descendants of these galaxies to examine their properties at the present day.

The layout of the paper is as follows.
Brief descriptions of the \Eagle\ simulations and the galaxy selection can be found in section \ref{sec:galselect}, where we also discuss the sensitivity of galaxy size to its definition. 
In section \ref{sec:normal} we consider the sizes of the full galaxy population.
We compare the sizes of simulated galaxies, and their evolution, to observational measurements in section \ref{sec:obs}, finding good agreement, particularly relative to previous simulation results (Appendix \ref{ap:comp}). 
In section \ref{sec:scatter} we determine the origin of the scatter in the \smr.
We then explore the evolution of individual galaxies in section \ref{sec:sizeevo} and the mass dependence of the size growth in section \ref{sec:mergers}.
In section \ref{sec:highz} we look at the evolution of high-redshift passive galaxies, with a particular focus on compact galaxies.
A summary of our main findings and concluding remarks are presented in section \ref{sec:summary}.

%% file: Simulations.tex
\section{Galaxies in the \Eagle\ Simulations}
\label{sec:galselect}

The production and selection of simulated galaxies are described below.  We begin with an overview of the primary \Eagle\ simulation in section \ref{sec:simulation}, followed by a discussion of the definition and selection of galaxies applied to the simulation in section \ref{sec:galdef}. In section \ref{sec:sizedef} we present our definition of galaxy size and discuss alternative choices.  Further details on alternative galaxy size definitions can be found in Appendix \ref{ap:sizes}.  Finally, the  galaxy classification applied in this study is introduced and discussed in section \ref{sec:ssfrcut}, with further details given in Appendix \ref{ap:cuts}.

\subsection{Simulation overview}
\label{sec:simulation}

For this study we use cosmological hydrodynamical simulations from the \Eagle\ project.  
We focus primarily on the largest of the simulations, \bigbox, a (100~cMpc)$^3$ simulation with baryonic particle masses of $1.2 \times 10^6$\msun\ which provides a large sample of galaxies resolved by at least 1000 star particles.
These galaxies span more than two orders of magnitude in stellar mass and are found in a diverse range of environments.

We also analyse two smaller, higher-resolution simulations, \highres\ and \highresref.
The high-resolution simulations are used for convergence tests presented in Appendix \ref{ap:res}.
The box sizes and resolutions of these simulations are summarised in Table \ref{table:res}.
The \highresref\ simulation has a factor of 8 (2) better mass (spatial) resolution and employs the same subgrid parameters as \bigbox; comparing these two simulations tests the strong convergence of the simulation (see \Schaye\ for definitions and discussion of the concepts of ``strong'' and ``weak'' convergence).
\highres\ also has a factor of 8 better mass resolution but was recalibrated to better reproduce the observed local Universe galaxy stellar mass function at this higher resolution.  This simulation enables examination of the weak convergence of the simulation.

The \Eagle\ simulation suite adopts a flat $\Lambda$CDM cosmogony with parameters inferred from the Planck data \citep{Planck13};  $\Omega_\Lambda = 0.693$, $\Omega_{\rm m} = 0.307$, $\Omega_{\rm b} = 0.048$, $\sigma_8 = 0.8288$, $n_{\rm s} = 0.9611$ and $H_0 = 67.77$ km s$^{-1}$ Mpc$^{-1}$. The \cite{Chabrier03} stellar initial mass function (IMF) is adopted. 
Galaxy stellar masses are computed by summing the mass of star particles located within
 a spherical aperture of 30 physical kiloparsecs (pkpc) centred on the potential minimum of the galaxy, as used by \Schaye. Star formation rates are computed within the same aperture. 
Galaxy sizes are quoted in physical units (e.g. pkpc), unless stated otherwise. 

A complete description of the \Eagle\ code and subgrid physics was presented by \Schaye; here we provide a concise overview.
Initial conditions at $z=127$ were generated using 2$^{\rm nd}$ order Lagrangian perturbation theory \citep{Jenkins10}.
The initial conditions were evolved using a parallel N-body smooth particle hydrodynamics (SPH) code, a modified version of \Gadget-3 \citep[based on \Gadget-2, last described by][]{Springel05}.
The SPH implementation in \Eagle\ includes the \cite{Hopkins13} pressure-entropy formulation, a time-step limiter \citep{Durier12}, artificial viscosity and conduction \citep{Dehnen&Aly12} and the \cite{Wendland95} C2 kernel; this set of changes to the SPH is collectively referred to as \Anarchy\ \citep[][see also Appendix A of Schaye et al. 2015]{DallaVecchia14}.
\Anarchy\ alleviates the problems associated with standard SPH in modelling contact discontinuities and fluid instabilities.
\cite{Schaller15} show, however, that at the resolution of \Eagle\ the impact of the hydrodynamics solver on galaxy properties in simulations is small for $M_* \lesssim 10^{10.5}$\msun.

To model astrophysical processes that occur on scales below the resolution of the simulation, subgrid schemes are applied. 
Included are subgrid models for radiative cooling and photo-heating, star formation, stellar mass loss and metal enrichment, stellar feedback from massive stars,  black hole growth and feedback from active galactic nuclei (AGN).

The cooling and photo-heating model uses the implementation of \cite{Wiersma09a}.  Abundances of 11 elements are considered when computing radiative cooling rates, tabulated from \Cloudy\ \citep[version 07.02][]{Cloudy} assuming ionization equilibrium and exposure to the cosmic microwave background and the \cite{HaardtMadau01} model for the evolving UV/X-ray background.

Above a metallicity-dependent density threshold, star particles form stochastically, with the star formation rate determined from the gas pressure such that it reproduces the observed Kennicutt-Schmidt law \citep{Kennicutt98, Schaye08}.
Star particles represent simple stellar populations described by a Chabrier IMF.
The Jeans length for the cold phase of the interstellar medium (ISM) cannot be resolved by the simulations and consequently a pressure floor is imposed,corresponding to a polytropic equation of state ($P_{\rm eos} \propto \rho^{\gamma_{\rm eos}}$), enabling the multi-phase ISM to be treated as a single-phase fluid.

Stellar mass loss is based on the implementation of \cite{Wiersma09b} where mass lost from asymptotic giant branch stars, through winds of massive stars, and supernovae (both core collapse and type Ia), is returned to the ISM over the lifetime of the stellar population.

Prompt feedback associated with star formation is implemented following \cite{DallaVecchia12} where the thermal energy available from stars formed is stochastically distributed to neighbouring gas particles, without any preferential direction. 
Energy injection always results in a temperature increase of $\Delta T$~=~10$^{7.5}$~K; such a temperature jump is required to mitigate numerical radiative losses at the resolution of the \Eagle\ simulations \citep{DallaVecchia12}.
The probability of energy injection depends on the metallicity and density of the local environment in which the star particle formed and is calibrated to reproduce the observed local Universe galaxy stellar mass function and the \smr\ for disk galaxies, where the galaxy sizes are a calibration diagnostic that is necessary to break degeneracies between different models that reproduce the galaxy stellar mass function \citep{Crain15}.
See \Schaye\ and \cite{Crain15} for a full description and motivation of the energy scaling.

The gas particle at the minimum of the potential of a halo is converted to a seed black hole of $10^{5}$\hmsun\ when a halo first reaches a mass of $10^{10}$\hmsun\ \citep{Springel05b}.
The sub-grid black hole accretes matter based on the modified Bondi-Hoyle model of \cite{Rosas-Guevara14}, adapted as described by \Schaye, which reduces the accretion rate for high angular momentum gas.
Of the matter accreted, a fraction of 0.015 of the rest-mass energy is returned to the surrounding medium in the form of energy.  
This feedback from AGN is implemented thermally, as for the stellar feedback, but with a temperature jump of $\Delta T~=~10^{8.5}$~K for Ref models and $\Delta T~=~10^9$~K for the Recal model.

\begin{table}
\centering
\caption{Simulation box size, particle number, initial baryonic and dark matter particle masses and maximum gravitational force softening for \bigbox, \highres\ and \highresref\ simulations.}
\smallskip
 \begin{minipage}{8.5cm}
\centering
\tabcolsep=0.11cm
  \begin{tabular}{|l |r | r| c| c| c|}
   \hline
   Simulation&Box size&$N_{\rm part}$&$m_{\rm g}$&$m_{\rm dm}$&$\epsilon$\footnote{For the redshift range covered in this paper the gravitational softening is fixed in the listed physical coordinates. At redshifts above 2.8 it is fixed in comoving coordinates.} \\
   \hline
    &[cMpc]& &[M$_\odot$]&[M$_\odot$]&[\kpc]\\
   \hline
   L100N1504&100&$2\times1504^3$&1.81$\times10^6$&9.70$\times10^6$&0.70 \\
   L025N0752&25&$2\times752^3$&2.26$\times10^5$&1.21$\times10^6$&0.35 \\
   \hline
  \end{tabular}  \par 
  \end{minipage}
\label{table:res}
\end{table} 

\subsection{Sample selection}
\label{sec:galdef}

Galaxy identification is carried out as described by \Schaye.
The friends-of-friends method with a linking length of 0.2 identifies overdensities that we refer to as halos.
Self-bound substructures within halos are identified using \Subfind\ \citep{Springel01,Dolag09}.
We refer to the stellar component of the subhalos as the galaxies in the simulation. 
A central galaxy corresponds to the most massive subhalo within a halo.

To enable comparison to the observational study of \vdw, we bin galaxies in redshift intervals of $\Delta z =0.5$. 
In the four redshift bins spanning $0\leq z\leq2$, we combine between 3 and 6 simulation outputs.
We include galaxies with stellar masses $M_* > 10^9$\msun, to ensure that the masses, sizes and star formation rates of galaxies are well sampled (see \Schaye\ and \Furlong\ for convergence studies). 
We limit the sample to galaxies with $z \leq 2$, as the typical size of galaxies at higher redshifts becomes comparable to the spatial resolution limit of the simulation, particularly for passive galaxies.
The resolution limit is set by the minimum of the gravitational force softening and the Jeans length ($\sim 2{\rm pkpc} f_{\rm gas} (n_{\rm H}/0.1{\rm cm}^{-3})^{-0.5}(T/10^4{\rm K})^{0.5}$, where $f_{\rm gas}$ is the gas fraction,  $n_{\rm H}$ is the hydrogen number density and $T$ is the temperature).  
The polytropic equation of state is implemented to avoid artificial fragmentation, thus suppressing fragmentation below the Jeans length.
This value, however, varies depending on the gas fraction and the density at which stars in the galaxy formed.

The sample contains 208,953 galaxies over the redshift range 0 to 2, of which 182,481 are actively forming stars and 26,472 are passive as defined in section \ref{sec:ssfrcut}. 
Further details of the galaxy sample can be found in Table \ref{table:numbers}.

\begin{table}
\centering
\caption{The number of galaxies with $M_* > 10^9$\msun\ in different redshift bins between $z_{\rm low}$ and $z_{\rm high}$, including the division into active and passive galaxies based on specific star formation rates. The SSFR separating active and passive galaxies varies with redshift and is given by log$_{10}$(SSFR$_{\rm lim}(z)/$Gyr$^{-1}) = 0.5z - 2$, $z \in [0, 2]$.}
\smallskip
 \begin{minipage}{8.5cm}
\centering
  \begin{tabular}{|c |c | r| c| c|}
   \hline
$z_{\rm low}$ & $z_{\rm high}$ & Total & Active& Passive  \\
& & & Fraction & Fraction \\
   \hline
0.0 &  0.5 & 80816  &  0.78 &  0.22 \\
0.5  & 1.0  &65682  &  0.90 &  0.10 \\
1.0  & 1.5  &34669  &  0.95 &  0.05 \\
1.5  & 2.0  &27786  &  0.97 &  0.03 \\
0.0  & 2.0  &208953 &  0.87 &  0.13 \\
   \hline
  \end{tabular}  \par 
  \end{minipage}
\label{table:numbers}
\end{table} 

We use galaxy merger trees to trace the evolution of individual galaxies. The merger trees are built by tracing the most bound particles of each subhalo between consecutive snapshots to determine the progenitor galaxies of all $z=0$ galaxies. 
The main progenitor branch is that with the largest branch mass, as defined by \cite{DeLucia07}.
This method is similar to that of the D-halo merger trees described by \cite{Jiang14}. 
A full description of the trees is presented by \cite{Qu15}, with their public release\footnote{www.eaglesim.org/database.html} discussed by \cite{McAlpine15}.

\subsection{Galaxy sizes}
\label{sec:sizedef}
While the abundance, range and quality of the observational data available to measure galaxy sizes are continuously improving, the recovered sizes from the data typically depend, among other things, on the wavelength of the observations \citep[e.g.][\vdw]{Kelvin12, Lange15}, the assumed fitting profile \citep[e.g.][]{Head15} and the surface brightness of galaxies.
While it is possible to ``observe'' simulated galaxies in a similar way to real galaxies \citep{Torrey15, Trayford15b} and use these observations to measure galaxy sizes that can be compared to a particular data set, in this study we are interested in the physical growth and evolution of galaxies, and in making predictions that can be compared with a variety of data sets. 
Hence, we use the physical half-stellar mass radius, $R_{50}$, as our definition of galaxy size, thus limiting systematic effects related to observations.

The \hmr\ is based on the total mass of all gravitationally bound stellar particles within a spherical aperture of radius 100~\kpc\ about the minimum of the gravitational potential of a galaxy. 
This definition of the \hmr\ of simulated galaxies is not directly dependent on the galaxy's shape or surface brightness.
We apply the 100~\kpc\ aperture to ensure that star particles located far out in the halo, but which are assigned to the galaxy by the subhalo finder, are excluded from the calculation. 
The aperture only affects the most massive galaxies in the simulation, with $M_* > 10^{11}$\msun, as shown in Appendix \ref{ap:sizes}.

Note that the size definition here differs from that used by \Schaye, where \sersic\ fits to the projected stellar mass profile within 30 \kpc\ apertures were used to measure the half-mass radius.
As mentioned above, in this study we aim to avoid assumptions about the profile of the galaxies.
A comparison between the sizes used by \Schaye\ and here can also be found in Appendix \ref{ap:sizes}.

\subsection{Galaxy classification}
\label{sec:ssfrcut}
We distinguish between galaxies that are actively forming stars and those that are passive.  
The division between active and passive galaxies may be based on a cut in colour or specific star formation rate, $\dot{M}_*/{M}_*$ (SSFR). 
Following \Furlong, we apply a SSFR cut that evolves with redshift, chosen to lie approximately 1 decade below the observed main sequence of star formation at each redshift.
We set the SSFR limit as log$_{10}$(SSFR$_{\rm lim}(z)/$Gyr$^{-1}) = 0.5z - 2$, $z \in [0, 2]$.
It was shown by \cite{Furlong15} that the fraction of passive galaxies in the simulation is broadly consistent with observations over this redshift range.
The impact of variations in the definition of active and passive galaxies on galaxy sizes is presented in Appendix \ref{ap:cuts}, while \cite{Trayford15b} show that the percentage of galaxies identified as passive is somewhat sensitive to this definition.

%% file: Results.tex
\section{Size evolution of the general population}
\label{sec:normal}

We begin in section \ref{sec:obs} with a comparison of the predicted \smr\ with observations from $0\leq z \leq2$. 
The scatter in the relation is considered in section \ref{sec:scatter}.
Having demonstrated in these sections that \Eagle\ reproduces the observed trends for galaxy sizes, we explore the size evolution of individual galaxies in section \ref{sec:sizeevo} and the variation in size evolution with mass in section \ref{sec:mergers}.

\subsection{The redshift dependence of the \smr}
\label{sec:obs}
\begin{figure*}
  \centering
  \includegraphics[width=1.0\textwidth]{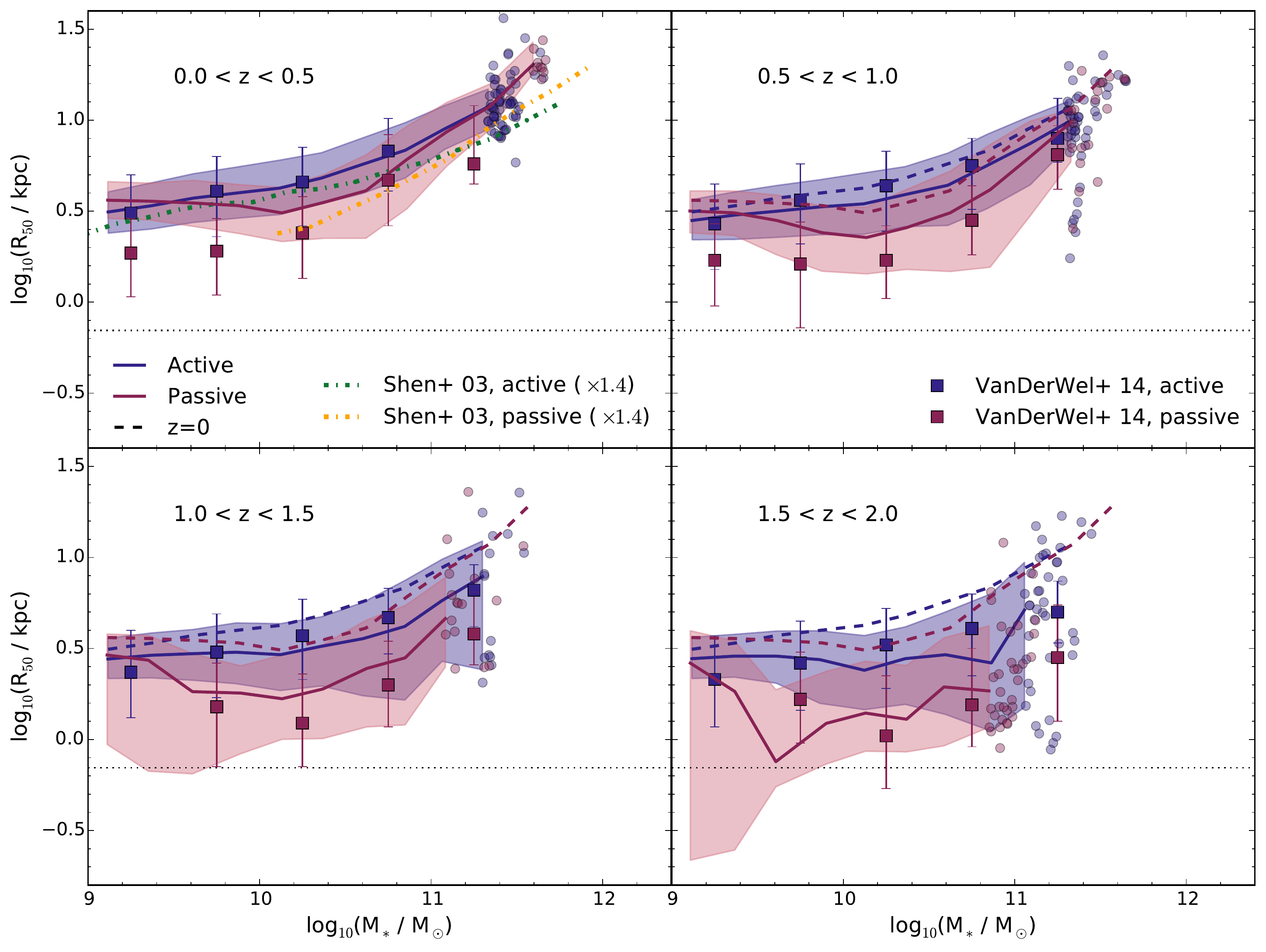}
  \caption{
Galaxy size -- mass relation in the \bigbox\ \Eagle\ simulation compared to observations.
Panels show different redshifts, from 0 to 2.	
Active and passive galaxies are shown in blue and red, respectively.
The median relations from the simulation are shown by solid curves, while the shaded regions enclose the 16\tht\ to 84\tht\ percentiles.
Bins of 0.2 dex in stellar mass are used.
Individual points are shown for bins containing fewer than 10 galaxies.
For reference, the dashed curves repeat the \smr\ for active and passive galaxies from the lowest redshift bin.
The horizontal dotted line corresponds to the gravitational force softening in the simulation, which is fixed at 0.7 \kpc\ over this redshift range.
Observational measurements presented by \citet{vanderWel14} are shown as squares with error bars representing the median and 1-$\sigma$ scatter.
Measurements from \citet{Shen03} are shown in the $0 < z < 0.5$ redshift bin, multiplied by 1.4 to account for the use of circularised radii (see text).
The observed trends with stellar mass, redshift and galaxy type are all broadly reproduced by the simulation, as well as the extent of the scatter.
The median relations found in the simulation agree with those inferred by \citet{vanderWel14} to within 0.1 dex for active galaxies and 0.2 dex for passive galaxies.
Passive galaxies at $M_*<10^{9.5}$\msun\ are larger than those observed, however this is a resolution effect (see text).
}
 \label{fig:size_mass}
\end{figure*}

The galaxy \smr\ is shown in Fig.~\ref{fig:size_mass}, with different panels showing the relation at different redshifts. 
In the simulation, galaxy sizes decrease with increasing redshift at fixed stellar mass (dashed curves repeat the relations at $0< z <0.5$ for reference). We restrict the comparison to $z < 2$, where the sizes of simulated galaxies typically lie above the gravitational force softening used in the simulation (indicated by horizontal dotted lines).
At $z < 1$ size increases with stellar mass, with a larger increase in size per stellar mass increment above $\sim 10^{10.5}$\msun. 
In the interval $1 \leq z \leq2$, galaxy sizes show a weaker trend with stellar mass. 

In each panel, the simulated galaxy sample is separated into active and passive galaxies. Passive galaxies are typically smaller than active galaxies by $>0.1$ dex at all redshifts for stellar masses between $10^{9.5}$ and $10^{11}$\msun. At higher masses (which are only sampled at $z < 1$ by the simulation) the \smr s of active and passive galaxies converge. 
At $M_* < 10^{9.5}$\msun\ the \smr s of both galaxy types are also similar, but this is a consequence of limited resolution. 
It was shown by \cite{Schaye15} and \cite{Furlong15} that for $M_* < 10^{9.5}$\msun\ the star formation rates of galaxies are poorly sampled compared to the higher-resolution L025N0752 simulations.
As a result, in this mass regime the classification into passive and active galaxies is not robust.
Furthermore, the Jeans length corresponding to the equation of state that is imposed on the unresolved interstellar medium may be a limiting factor for the galaxy sizes in this regime (for $R_{50} \lesssim 1$ \kpc).

To compare the \smr s of simulated galaxies with observational constraints, we also plot the measurements of \vdw\ in Fig. \ref{fig:size_mass}. 
This enables us to compare with a set of uniformly-analysed observational measurements over the full redshift range.
The effective radius of galaxies in the \vdw\ sample is defined as the major axis of the ellipse containing half the total flux associated with a single component \sersic\ fit applied to the 2-D light profile of the galaxy.
The effective radius is converted to a rest-frame wavelength of 5000\AA\ by empirically establishing the dependence of the radius on wavelength.
We supplement the \vdw\ measurements with $z=0$ counterparts presented by \cite{Shen03}, since the limited volume of the \Candels\ survey used by \vdw\ precludes it from adequately sampling galaxies with $M_* > 10^{11}$\msun\ in the lowest redshift bin.
\shen\ infer the \smr\ from SDSS \citep{Stoughton02} over the redshift range 0 to 0.3, where galaxy type is defined using concentration, \sersic\ index and photometric colours.  
Their larger sample in this redshift bin extends the observations of the \smr\ to $M_* \sim 10^{12}$\msun.
Note, however, that the circularised radii\footnote{$R_{\rm circ}=\sqrt{\frac{b}{a}}R_{\rm maj}$, where $b$ and $a$ are the minor and major axes and $R_{\rm maj}$ is the major axis size.} used by \shen\ are typically a factor of 1.4 smaller than uncircularised radii, or the major axis radii, used by \vdw, as discussed by \cite{Dutton11}. 
To account for this difference, the \shen\ sizes are multiplied by a factor of 1.4 in Fig. \ref{fig:size_mass}.  For both active and passive galaxies the measurements of \shen\ and \vdw\ are similar at $M_* < 10^{10.5}$\msun, with active galaxies typically being larger than passive types of the same stellar mass for $M_* < 10^{11}$\msun.
At the higher masses probed only by \shen, the sizes of both types overlap.
At higher redshifts the normalisation of the observed \smr\ decreases while the offset in size between the active and passive galaxies remains.

The trends seen in the simulation, namely an increase in galaxy size with stellar mass, an offset in size between active and passive galaxies and an increasing size with decreasing redshift, are all consistent with the observed trends. 
Although the normalisation of the \smr\ of local galaxy disk sizes was used to calibrate the simulations, it is remarkable that the relation for active galaxies traces that of the \vdw\ observations within $\sim 0.1$ dex for all stellar masses shown up to $z = 2$. 
In the lowest redshift bin at $M_{*} > 10^{11}$\msun\ however, where no \Candels\ measurements are available, there is a larger offset of $\sim0.2$ dex with respect to the \cite{Shen03} data, although we note that this may be the result of the relatively bright surface brightness limit of 
the SDSS survey (which can be mimicked by applying a different aperture to the simulated galaxies, the effect of which is demonstrated in Appendix \ref{ap:sizes}). 
The sizes of passive galaxies are consistent with the observational measurements to within 0.1 -- 0.2 dex, with the largest discrepancies at $M_*<10^{9.5}$\msun, where adequate resolution renders the definition of active and passive galaxies ambiguous.
A comparison with the sizes of simulated galaxies from the \Illustrius\ simulation \citep{Illustrius}, which are found to be a factor of two larger than in \Eagle\ and observations, is presented in Appendix \ref{ap:comp}.


In the following sections, we capitalise on the correspondence between \Eagle\ galaxies and the observational measurements to interpret the 
evolution of galaxy sizes in physical terms. 

\begin{figure}
  \centering
  \includegraphics[width=0.5\textwidth]{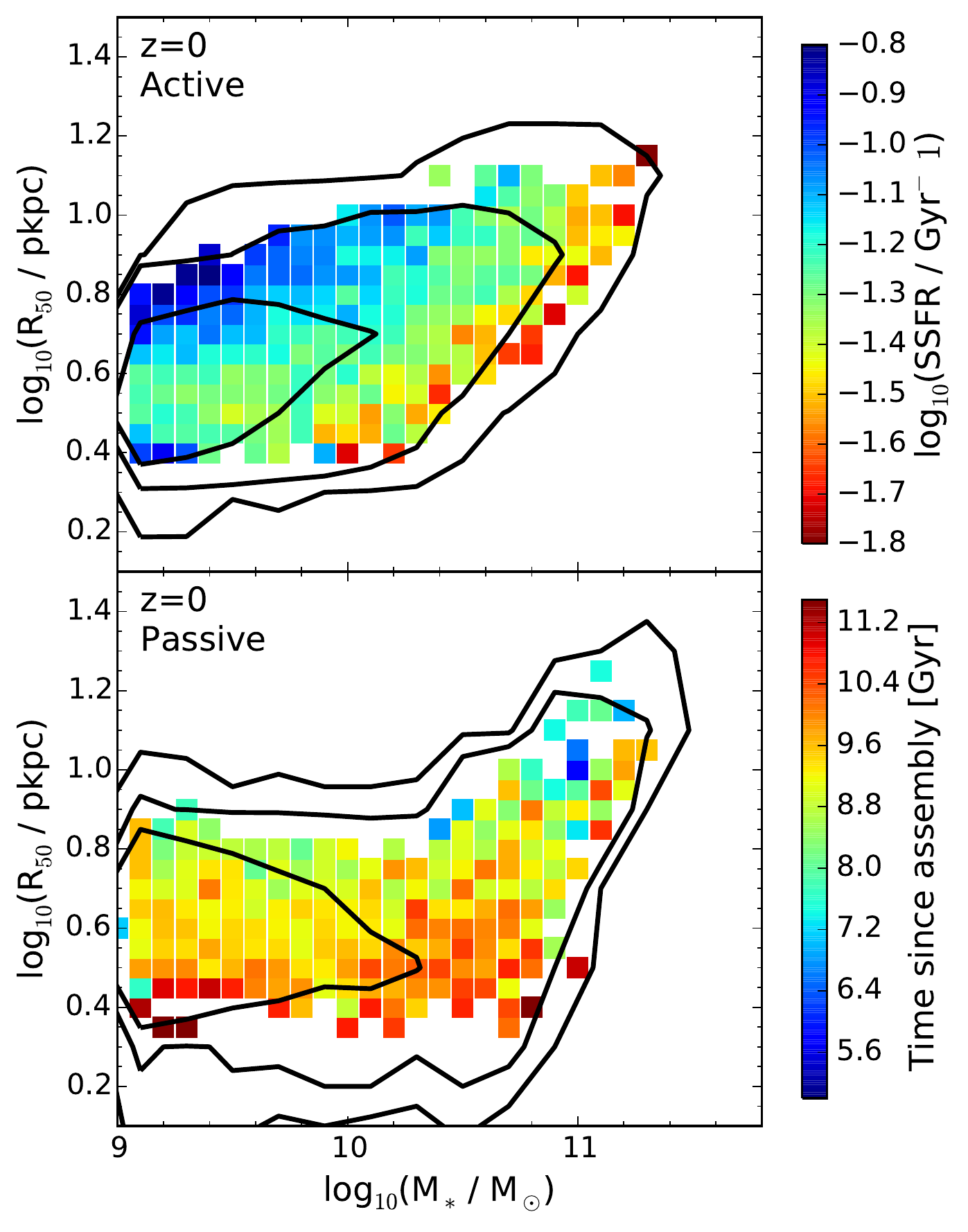}
  \caption{
The $z=0$ \smr s for active and passive galaxies are shown in the top and bottom panels respectively.
Black contours indicate the 1, 2 and 3 sigma levels.
The galaxy sizes and stellar masses are binned, with each point coloured by the median SSFR for active galaxies (upper panel) or the time since the main progenitor reached half its present-day stellar mass for passive galaxies (lower panel).
There is a clear correlation between the scatter in the \smr\ and these properties. 
	}
 \label{fig:scatter_1}
\end{figure}

\subsection{The scatter in the $z=0$ \smr}
\label{sec:scatter}
The scatter in the \smr s is quantified Fig. \ref{fig:size_mass},
with an overlap in sizes between active and passive galaxies. The scatter in the observational and simulated data sets is comparable. In this section, we explore potential origins of this scatter through its correlation with other galaxy properties. 

In Fig. \ref{fig:scatter_1} the \smr s at $z=0$ for active and passive galaxies are shown in the top and bottom panels respectively, with contours showing the number density of galaxies in the size -- mass plane.
In the top panel, the underlying colour image shows the median SSFR of galaxies in bins of $M_*$ and $R_{50}$. 
The figure shows a clear trend: at a fixed stellar mass, smaller galaxies have lower SSFRs. 


For passive galaxies, 
the present-day SSFR reflects sporadic low-mass accretion and no clear correlation between the SSFR and galaxy sizes is evident (not shown). A better measure of the past histories of passive galaxies is the time since they assembled, $t_{\rm lb, \rm assemble}$, defined as the lookback time at which 50\% of the stellar mass of the $z=0$ galaxy has assembled into a single progenitor galaxy.
The bottom panel of Fig. \ref{fig:scatter_1} is coloured by the median value of $t_{\rm lb,\rm assemble}$ in each bin.
At a given stellar mass, smaller galaxies typically assembled earlier.
A similar trend with $t_{\rm lb, \rm assemble}$ is seen for active galaxies (not shown).
 
The assembly time of a galaxy reflects the assembly history of the main progenitor, and accounts for growth due to both mergers
and in situ star formation. We have also examined the trends in the scatter with the time since formation, $t_{\rm lb, \rm form}$, defined as the lookback time at which 50\% of the stellar mass in all of a galaxy's progenitors has formed, thus eliminating the contribution of growth through mergers.  
For passive galaxies with $M_* < 10^{10.5}$\msun\ similar trends are seen for $t_{\rm lb, \rm form}$ and $t_{\rm lb, \rm assemble}$, but at higher masses galaxy sizes show no clear trend with $t_{\rm lb, \rm form}$, with all massive passive galaxies, irrespective of galaxy size, forming at similar times.
For these high-mass, passive galaxies, the mergers that the galaxy has undergone must therefore play a role in the galaxy's size evolution.
We will further examine the impact of mergers in section \ref{sec:mergers}.

\subsection{The size evolution of individual galaxies}
\label{sec:sizeevo}
In this section we contrast the evolution of the size of individual galaxies with the redshift dependence of the \smr\ for the ensemble population. 
Na{\"i}vely one might expect these to be equivalent.
This is, however, not the case since individual galaxies evolve in mass, can switch between active and passive states, and may merge.
Note that observational estimates of the size evolution of galaxies have been carried out using more sophisticated techniques such as number density matching \citep[e.g.][]{vanDokkum10} which aims to account for the evolution of galaxy masses.
However, accounting for the transition between active and passive states is more difficult.

\begin{figure*}
  \centering
\minipage{1.0\textwidth}
  \includegraphics[width=1.0\textwidth]{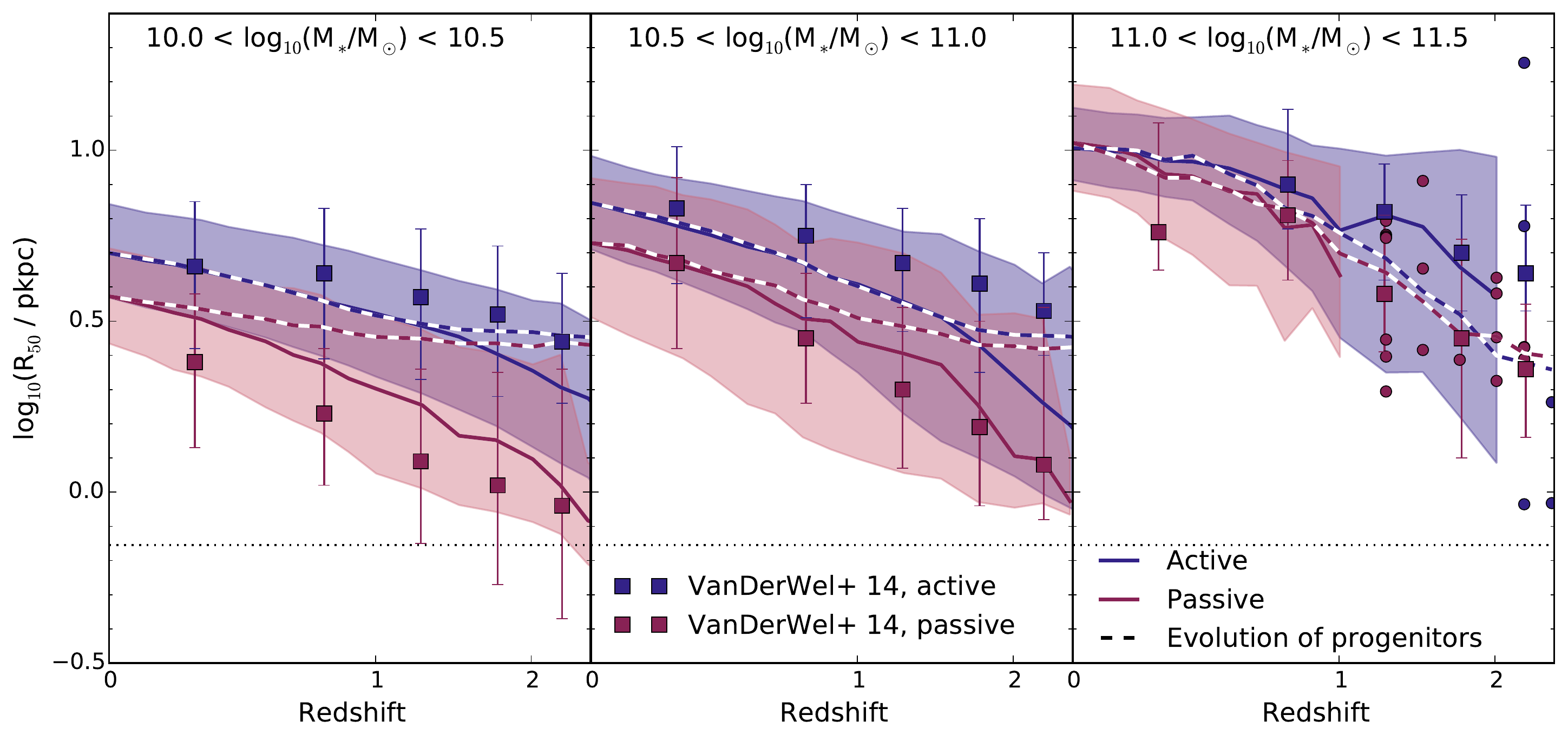}
\endminipage\hfill
  \caption{
The evolution of the \smr\ with redshift for active and passive galaxies in 3 stellar mass bins.
Active and passive galaxies are shown in blue and red respectively.
The median relations from the simulation are shown by solid curves, while the shaded regions enclose the 16\tht\ to 84\tht\ percentiles.
Individual points are shown for bins containing fewer than 10 galaxies.
Dashed lines show the median evolution of individual galaxies selected at redshift 0, the panel in which they appear is based on their $z=0$ masses and the active / passive classification is based on their $z=0$ SSFR.
The evolution of individual galaxies is determined from the merger trees.
Note that at $z>0$ the galaxies used to compute the dashed curves may have a stellar mass that falls outside the range indicated in each panel and that their classification into active / passive may not correspond to the colour of the line.
The horizontal dotted line corresponds to the gravitational force softening in the simulation.
Measurements of the median and 1-$\sigma$ scatter from \citet{vanderWel14} are shown as squares with error bars. 
The redshift dependence of the \smr\ for active galaxies is similar to the size evolution of individual active galaxies.
However, this is not true for passive galaxies, owing to the evolving number density of passive galaxies.
	}
 \label{fig:size_evo}
\end{figure*}

\begin{figure}
  \centering
  \includegraphics[width=0.5\textwidth]{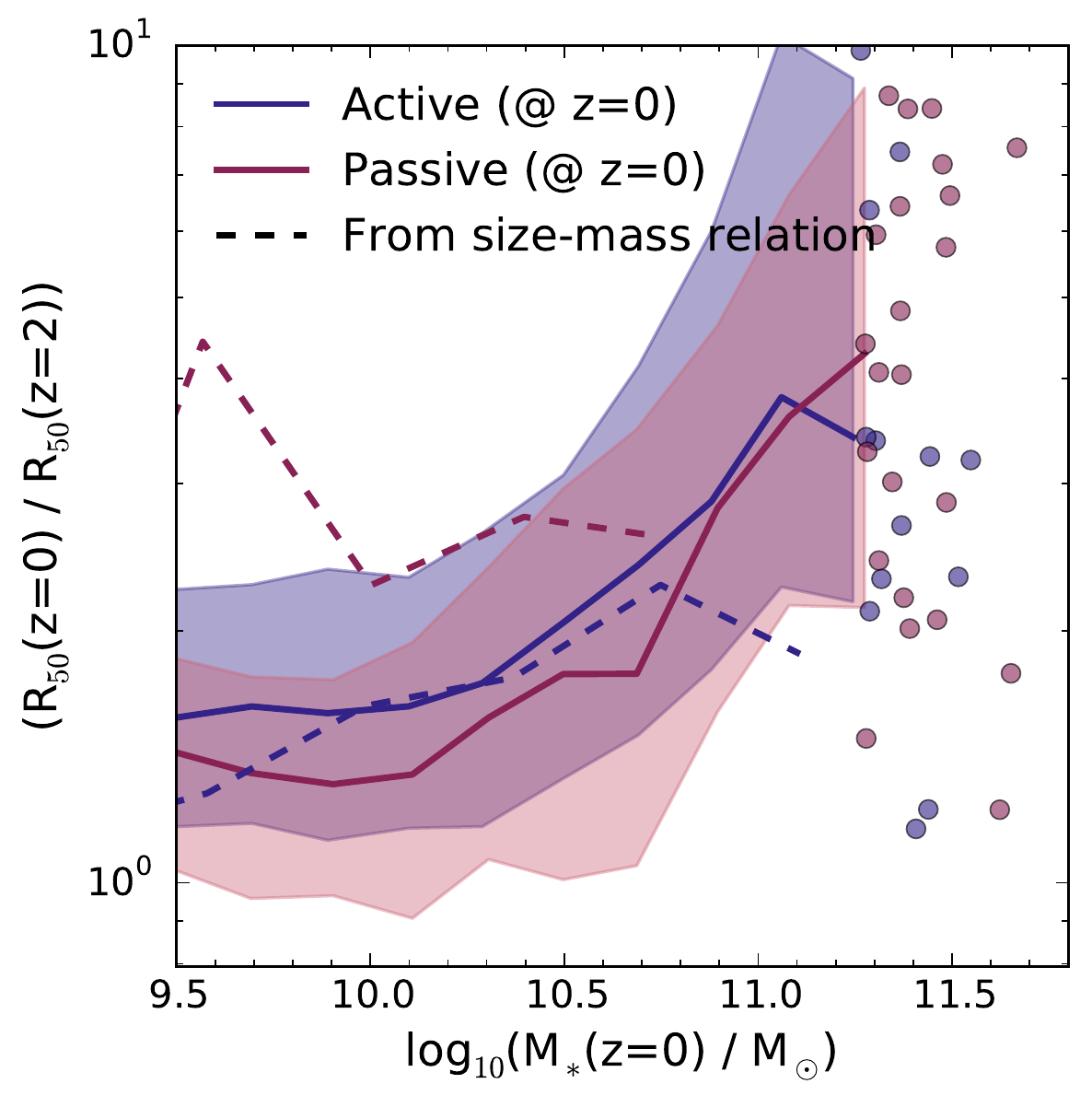}
  \caption{
The size growth, $R_{50}(z=0)/R_{50}(z=2)$, from redshift 2 to 0 as a function of stellar mass.
Solid lines show the median size growth from $z=2$ to $z=0$ for individual galaxies selected based on their $z=0$ stellar mass and SSFR.
Galaxies that are active and passive at $z=0$ are shown in blue and red.
The size growth of these galaxies is determined based on their main progenitors at $z=2$.
Individual galaxies are shown when there are fewer than 10 galaxies per stellar mass bin.
The shaded regions enclose the 16\tht\ to 84\tht\ percentiles.
Dashed lines show the difference in the medians of the \smr s at $z = 2$ and $z=0$, based on the data presented in Fig. \ref{fig:size_mass}.
Blue and red colours show the median ratios of the active and passive galaxies respectively.
Note that for the dashed curves, the stellar mass plotted along the x-axis applies at both $z=2$ and $z=0$.
The size growth of individual galaxies depends strongly on $z=0$ stellar mass but only weakly on $z=0$ galaxy type. 
The growth determined from the redshift dependence of the \smr\ for passive galaxies overestimates the median size growth of individual galaxies by factors of a few.
	}
 \label{fig:deltaR_m}
\end{figure}

In Fig. \ref{fig:size_evo}, the solid curves and shaded regions indicate the normalisation and scatter of the \smr\ for three bins of stellar mass at each simulation output. Note that the galaxies that fall in a particular bin can be different at each redshift due to the stellar mass evolution of individual galaxies.
Comparing the sizes of galaxies over the redshift interval $z=2$ to 0, the sizes of galaxies increase for all stellar masses, as seen in Fig. \ref{fig:size_mass}.
The trends are similar for active and passive galaxy \smr s, but the very small sizes of the passive galaxy median \smr\ at high redshifts means that the factor by which the passive sequence increases in size is somewhat greater than for active galaxies (as quantified in Fig. \ref{fig:deltaR_m}, dashed curves).  
For comparison, observational data from \vdw\ is represented by points with error bars; this re-emphasises the relatively good agreement between \Eagle\ and observations discussed in section \ref{sec:obs}.

We now compare these trends to the typical size evolution of {\it individual} galaxies.
We select a sample of galaxies at $z=0$ and compare the evolution of their main progenitors' sizes to the redshift dependence of the \smr.
We select galaxies in three $z=0$ stellar mass bins, and separate the population into active and passive types on the basis of their $z=0$ SSFRs. 
We then trace the galaxies back in time using the merger trees described in section \ref{sec:galdef}, identifying a galaxy's main progenitor at each previous simulation output.
We combine the galaxies in each $z=0$ sample and over plot their median size evolution using dashed lines in Fig. \ref{fig:size_evo}. 

It is important to note that galaxies which are passive at $z=0$ will not necessarily be passive at higher redshifts (and similarly for active galaxies - see section \ref{sec:highz}). 
In fact we see that the sizes of the samples that are active and passive at $z=0$ become similar at higher redshift.  
Indeed, the comoving number density of passive galaxies at $z=2$ is less than $10\%$ of that at $z=0$ and most of the galaxies that are passive at $z=0$ were active at high redshift.
As a consequence, the sizes of the passive $z=0$ progenitors lie above the \smr\ of \highz\ passive galaxies, since passive galaxies are typically smaller than active galaxies at a given stellar mass. 
The evolution of the \smr\ for passive galaxies is thus unlikely to reflect the evolution of individual passive galaxies.
A similar conclusion was drawn by \cite{Carollo13} based on observational measurements.

Galaxies that are selected to lie in each of the stellar mass bins at $z=0$ will generally not be in the same mass bin at higher redshift.
For example, only 3\% of present-day $10^{10} < M_* < 10^{10.5}$\msun\ galaxies had progenitors with stellar masses in this range at $z=2$.
This results in an interesting feature in Fig. \ref{fig:size_evo} at $z\geq2$ for galaxies in the mass range $10^{10}$\msun\ $< M_* < 10^{11}$\msun\ where even active galaxies at this redshift are smaller by $\sim 0.2$ dex than the progenitors of galaxies that are active and have the same mass at $z=0$. 
The median \smr\ for active galaxies at $z=2$ has a shallow decrease in sizes with increasing stellar mass for $M_* > 10^{10}$\msun, with a similar trend at higher redshift, so lower-mass galaxies have larger sizes at $z \gtrsim 2$.
As the progenitors of the $z=0$ sample typically have lower masses than specified in each bin at $z=2$, their sizes fall $\sim 0.2$ dex above the median size \smr.

Not only are the sizes of the progenitors of galaxies that are passive at $z=0$ similar to their active counterparts, but active and passive $z=0$ galaxies experience broadly similar size growth.
In Fig. \ref{fig:deltaR_m}, we show the size of galaxies at $z=0$ relative to the size of their main progenitor at $z=2$ (which we refer to as the size growth).  
The solid lines show the median ratio of the $z=2$  and 0 sizes of galaxies selected based on their $z=0$ properties, while the dashed lines show size growth based on the \smr.
Note that for the solid lines the stellar mass shown on the x-axis applies to $z=0$.
For active galaxies the size growth determined na{\"i}vely from the redshift dependence of the \smr\ is similar to that based on individual galaxy histories, but the median size growth of individual passive galaxies is overestimated by the \smr\ by factors of a few.

\subsection{The dependence of size growth on stellar mass}
\label{sec:mergers}

\begin{figure}
  \centering
  \includegraphics[width=0.5\textwidth]{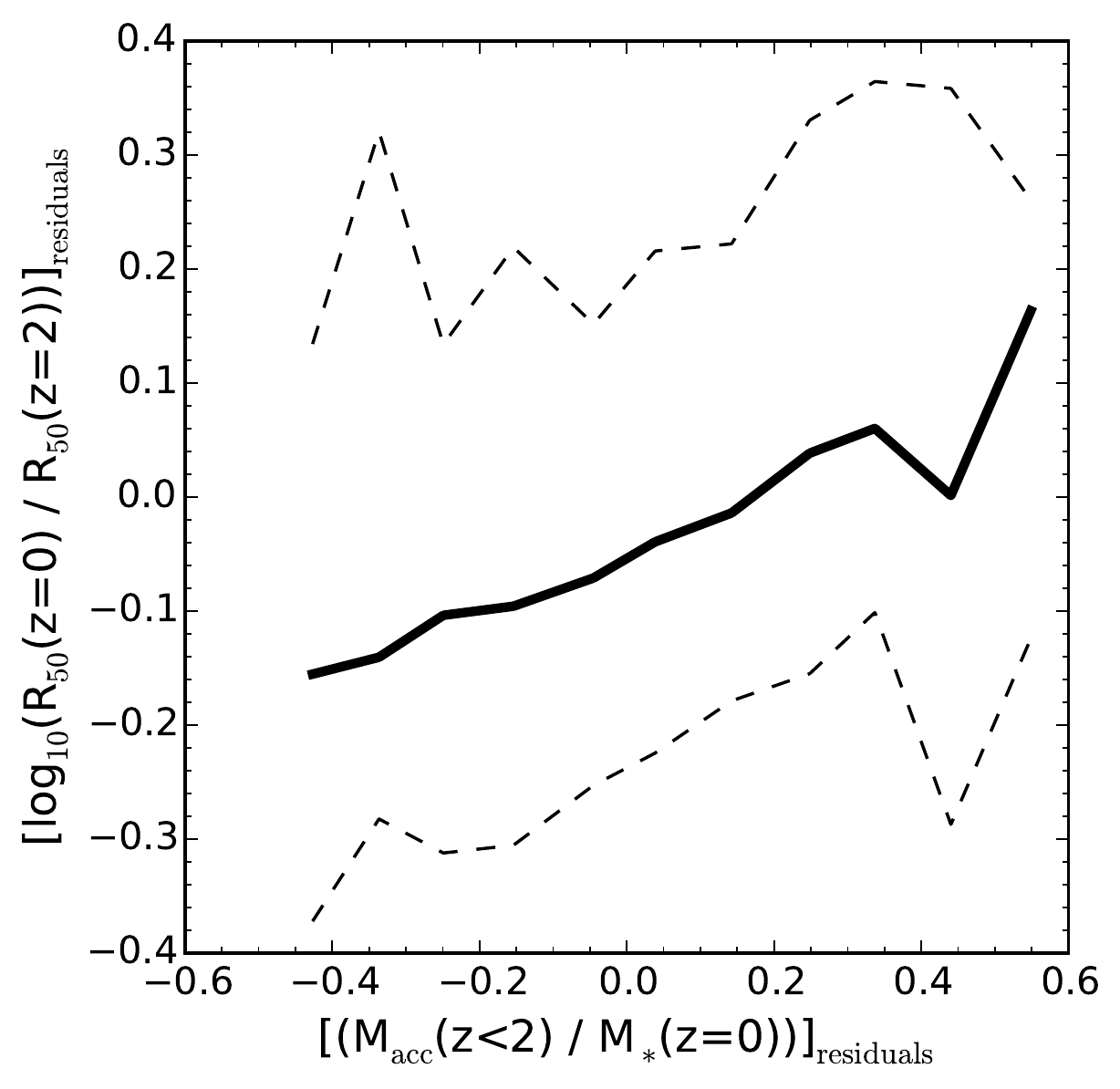}
  \caption{
Residuals of the relation between size growth since $z=2$ and stellar mass as a function of the residuals of the relation between accreted mass fraction and stellar mass. 
A positive correlation indicates that galaxies of a fixed stellar mass are
larger if they accreted a larger fraction of their mass.
The median relation is shown by the solid curve, the 16\tht\ to 84\tht\ percentiles are shown by the dashed curves.
The excess in size growth correlates well with an excess in accreted mass.
	}
 \label{fig:mergers}
\end{figure}

It is clear from Fig. \ref{fig:deltaR_m} that galaxy size growth from $z=2$ to $z=0$ depends strongly on $z=0$ stellar mass.
There is a strong trend with mass for both galaxy types, with the growth increasing from factors of $\sim$ 1 -- 2 for $M_* \sim 10^{9.5}$\msun\ up to median factors of 3 -- 4 for $M_*>10^{11}$\msun.
 This trend with $z=0$ mass is significantly more pronounced than any relation with a galaxy's $z=0$ active or passive classification, where the size growth differs only by $\sim 30\%$.
Given the weaker trend of the sizes with stellar mass at $z \sim 2$ relative to $z \sim 0$ in Fig. \ref{fig:size_mass}, a strong trend with stellar mass is anticipated.

A strong connection is expected between galaxy mergers and size growth \citep[e.g.][]{Cole00, Naab06, vanderWel09}.
As will be shown by \cite{Qu15}, the number of major and minor mergers experienced by galaxies in the \Eagle\ simulation is a strong function of stellar mass, consistent with findings inferred from observations of close projected pairs.
The correlation with $z=0$ stellar mass of both the size growth and the number of mergers since $z=2$ implies a correlation between size growth and mergers.  

To examine this correlation we consider the relation between the size growth and accreted mass fraction, where the accreted mass fraction is the ratio of stellar mass accreted since $z=2$ to the $z=0$ stellar mass. 
The accreted mass is defined as the contribution to the stellar mass of the galaxy from sources other than the main progenitor at a given time, where the stellar mass of the secondary branches is evaluated when these galaxies are at a radius of five times the half mass radius of the main branch progenitor.  
This definition ensures that any mass lost due to stripping, before the secondary branch coalesces with the main branch, are accounted for in the accreted mass \citep{Qu15}.

We show in Fig. \ref{fig:mergers} the residuals of the size growth - $M_*(z=0)$ relation as a function of the residuals of the accreted mass fraction - $M_*(z=0)$ relation.
We look at the residuals of these relations to negate the correlation of both the size growth and accreted fraction with stellar mass.  
In this way we test whether size growth and accreted mass fraction are inherently related.
Indeed, we see in Fig. \ref{fig:mergers} that this is the case: excess size growth increases with an excess in accreted mass fraction.
This implies that mergers directly boost size growth, beyond the stellar mass dependence of the merger rate.

In summary, we find that the growth in size of individual galaxies has a strong dependence on present-day stellar mass and a much weaker dependence on present-day galaxy classification. 
The merger history of a galaxy goes some of the way to explaining the dependence of size growth on $z=0$ stellar mass; galaxies that gain more of their mass from accretion experience larger size growth.

\section{Size evolution of passive high-redshift galaxies}
\label{sec:highz}

In section \ref{sec:obs}, we showed that (both in observations and in the simulation) passive galaxies at high redshift are typically small, $\sim 1-2$ \kpc. In the previous section, however, we saw that the main progenitors of present-day passive galaxies have  sizes that are similar to the progenitors of active galaxies. This raises an obvious question: what happened to the 
small passive galaxies that are seen at high redshift?

We compare the number density of passive compact galaxies to those observed in section \ref{sec:numdens}.
We look at the evolution of all \highz\ passive galaxies in section \ref{sec:highzevo} and focus on the evolution of the compact galaxy sample in section \ref{sec:compact}.

\subsection{The number density of compact galaxies}
\label{sec:numdens}
\begin{figure}
  \centering
  \includegraphics[width=0.5\textwidth]{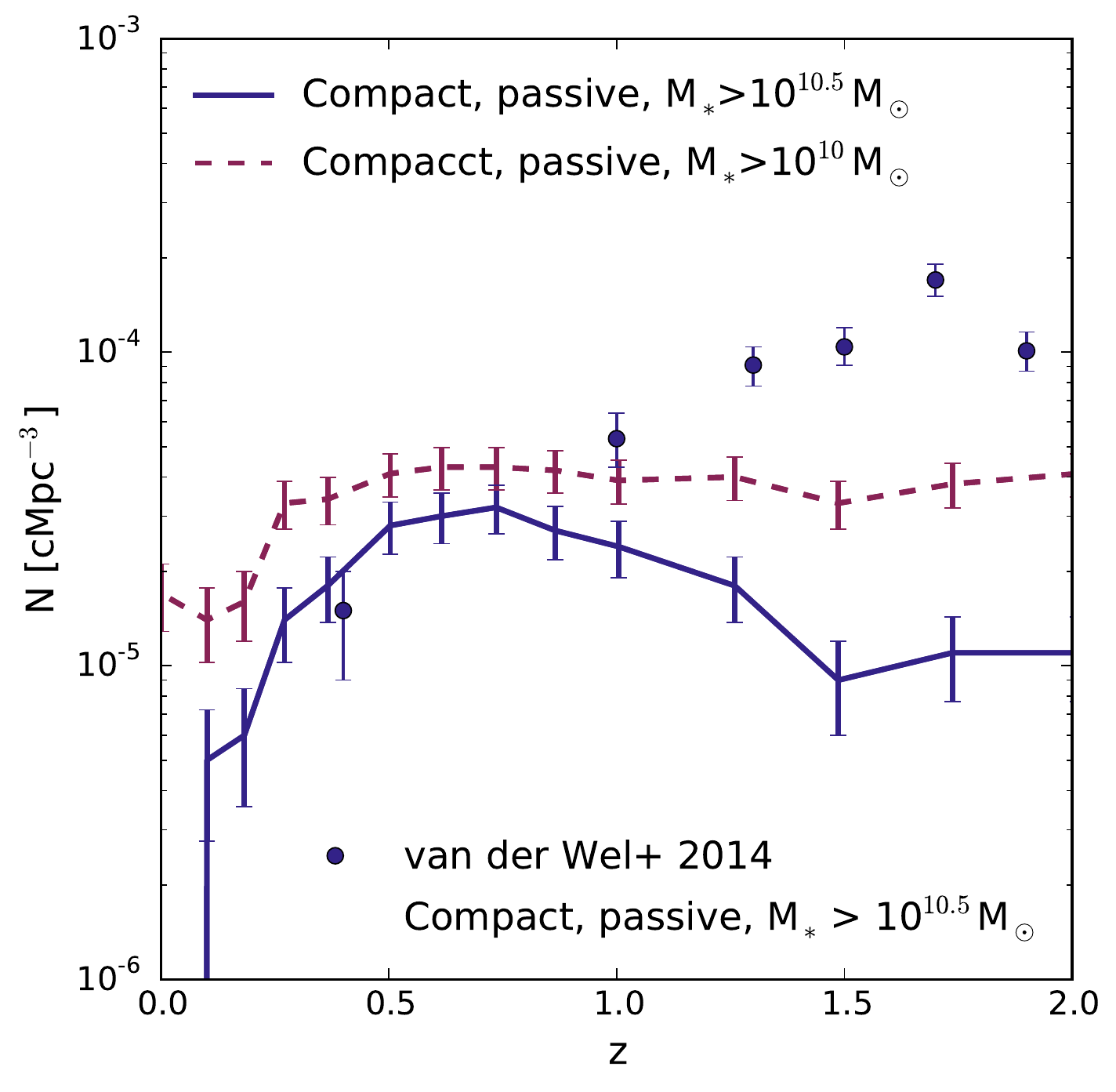}
  \caption{
The comoving number density of compact galaxies in the simulation as a function of redshift, where compact galaxies are those with $R_{50}/(M_*/10^{11}$\msun$)^{0.75} < 2.5$ \kpc, as defined by \citet{vanderWel14}.
The blue solid curve shows compact galaxies that are passive and have $M_* > 10^{10.5}$\msun, the red dashed curve lowers the mass limit to $M_* > 10^{10}$\msun.
Error bars on the simulation data indicate the Poisson error per redshift bin.
The observational measurements of \citet{vanderWel14} are shown as points with $1-\sigma$ error bars.
Many massive compact galaxies are identified in the simulation, however the number density falls
below that of the observations at $z\geq1$, which is at least in part due to the limited box size of the simulation.
	}
 \label{fig:numdens}
\end{figure}

Observational studies have detected massive compact red galaxies at high redshifts \citep[e.g.][]{Cimatti04, Daddi05, Trujillo06, vanDokkum08, Damjanov09}.
The comoving number density of such galaxies is observed to increase from $z \lesssim 3$, with evidence of a turn-over at $z < 1$ \citep[\vdw,][]{Damjanov15}\footnote{The existence of a turn-over depends on the definition of ``compact'', see \cite{Damjanov15} for a comparison of different selection criteria in the literature.}.
Several scenarios for the evolution of this population have been proposed: \cite{Trujillo06} suggest that the population must grow in size by factors of 3 to 6, for example through dry mergers, and that the descendants are today found on the \smr: \cite{vanderWel09} similarly suggest that growth through dry mergers, combined with the continuous emergence of passive galaxies, explains the evolution of such galaxies: \cite{Poggianti13} suggest that at most half of the population has evolved appreciably in size based on the luminosity-weighted ages of present-day compact galaxies: \cite{Graham15} conclude that the \highz\ compact galaxies may simply be the compact cores of local early-type disk galaxies today.

Recently, the theoretical study of \cite{Wellons15b} considered the evolution of 35 compact galaxies identified in the \Illustrius\ simulation, using a modified version of the \cite{Barro13} criterion for compactness.
Their key finding is that environment is an indicator of compact galaxy evolution, such that galaxies in denser environments undergo greater size evolution.
Note that while our results on compact galaxies are qualitatively comparable to \cite{Wellons15b}, they employ 
a compactness criterion based on that of \cite{Barro13} with modification ($\log_{10}(\Sigma_{1.5}) > 10^{11}$, where $\Sigma_{1.5} = (M_*/R_{50}^{1.5})/(\rm M_\odot$/\kpc$^{1.5}$)), which differs from the criteria of \vdw\ applied here.

We begin our analysis by defining our compact galaxy selection as those that are passive with $(R_{50}/{\rm pkpc})/(M_*/10^{11}$\msun$)^{0.75} < 2.5$, following \vdw.
The evolution of the number density of passive galaxies satisfying the compactness criteria in the simulation is shown in Fig. \ref{fig:numdens}.
Applying the same conditions as those applied to the \vdw\ sample, namely that galaxies are red/passive and have $M_*>10^{10.5}$\msun\ (solid curve), a reasonably flat number density evolution is recovered from $z=2$ to 0.5, with a decline for $z<0.5$.
Relaxing the mass constraint to include all galaxies with $M_*>10^{10}$\msun\ increases the number density of galaxies identified but does not change the overall evolutionary trend (dashed curve).
Compared to the observations of \vdw, also shown in Fig. \ref{fig:numdens}, the number density of compact galaxies in the simulation is too low, although there is a similar turn-over at low redshift.
This discrepancy is at least in part due to the limited box size of the simulation.
Indeed, comparing with a simulation in a smaller box of 50$^3$ cMpc$^3$ (with the same resolution and subgrid physics as in \bigbox) no compact galaxies are found with $M_* > 10^{10.5}$\msun\ implying an upper limit to the number density of $8 \times 10^{-6}$ cMpc$^{-3}$, which is much lower than the $\mathcal{O}(10^{-5}$) cMpc$^{-3}$ found at $0.5 \gtrsim z \gtrsim 2$ for our fiducial simulation.
Thus, it is likely that the environments in which compact galaxies form are undersampled in the \Eagle\ simulation volume.
Nonetheless, we are interested in the evolution of the identified \highz\ passive compact galaxies, even though the simulation may not account for all possible host environments of those observed.

\subsection{The destiny of \highz\ passive galaxies}
\label{sec:highzevo}

\begin{figure}
  \centering
  \includegraphics[width=0.5\textwidth]{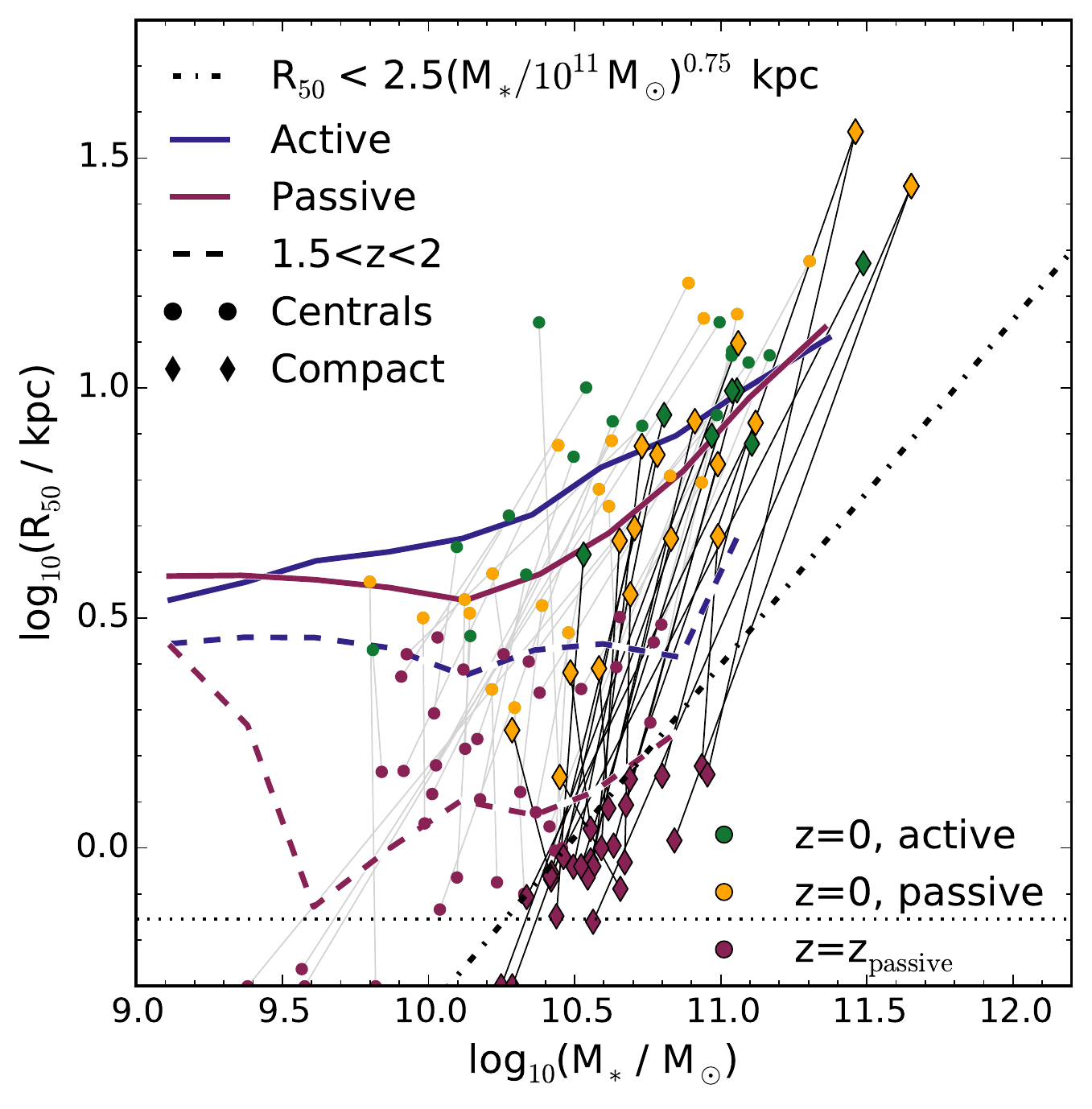}
  \caption{
The evolution of galaxies that are passive in the redshift range $1.5<z<2$ and centrals at $z=0$ in the size -- stellar mass plane from the redshift where they are first identified as passive, $z_{\rm passive}$, to $z=0$.
Points at $z_{\rm passive}$ are shown in red, while redshift 0 points are coloured by galaxy type, where orange indicates passive galaxies at $z=0$ and green indicates active at $z=0$.
A grey line connects the $z=z_{\rm passive}$ and $z=0$ points for each galaxy.
Any galaxy lying to the right of the dash-dotted line is defined as compact.
For the compact galaxies the evolution is highlighted using black lines and diamond symbols.
(Note that all compact galaxies are shown, not only those that are centrals at $z=0$.)
The median \smr s for active (blue) and passive (red) galaxies are shown at $z=0$ and for the range $1<z<2$ as solid and dashed lines respectively.
The horizontal dotted line corresponds to the simulation gravitational softening.
All compact galaxies increase in size by at least 0.2 dex from when they were first identified as passive, and similar for all centrals. 
None of the \highz\ compact galaxies remain compact to $z=0$.
Although all galaxies shown are passive in the interval $1.5<z<2$, not all are passive at $z=0$.
	}
 \label{fig:compact_evo}
\end{figure}

\begin{figure}
  \centering
  \includegraphics[width=0.5\textwidth]{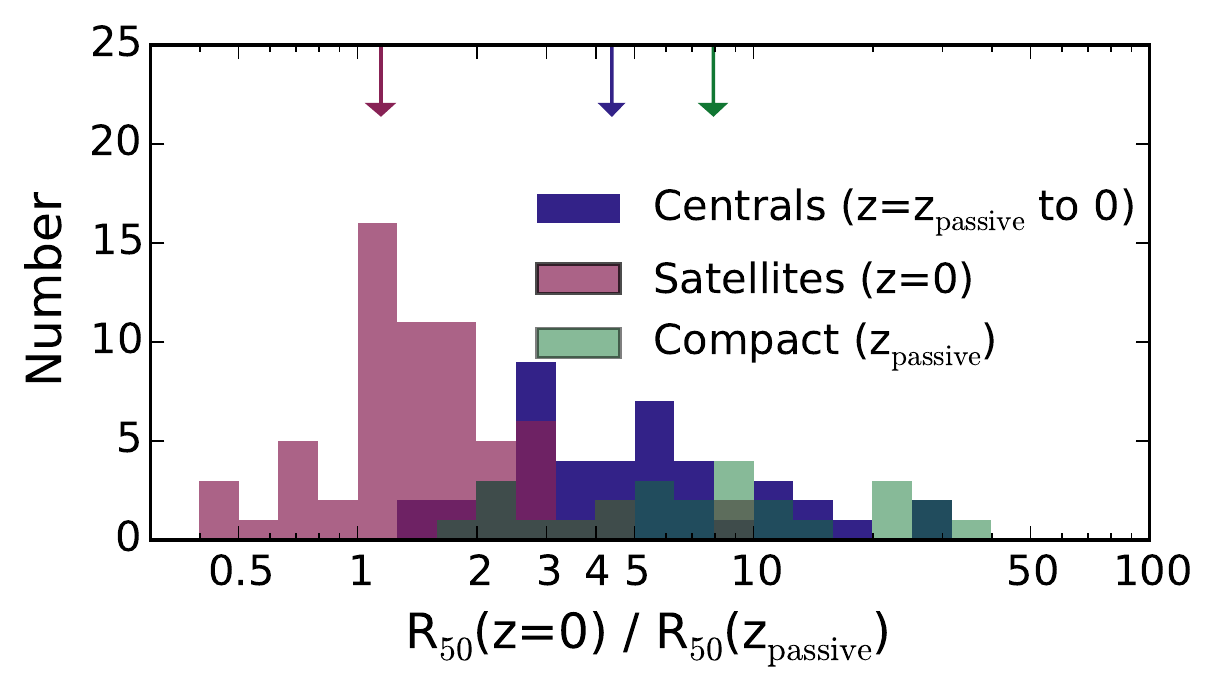}
  \caption{
The distribution of the relative size growth for \highz\ passive galaxies between the highest redshift in the range $1.5<z<2$ at which they are identified as passive and $z=0$.
Galaxies that remain centrals until $z=0$ are shown in blue, while $z=0$ satellites are shown in red.
The compact galaxies highlighted in Fig. \ref{fig:compact_evo} are shown in green, these include centrals, satellites and merged galaxies.
The median size growth of each sample is indicated by an arrow.
All galaxies show broad distributions in size growth, with several satellite galaxies decreasing in size.
	}
 \label{fig:gdist}
\end{figure}

As already noted, the typical evolution of all \highz\ passive galaxies cannot be understood from the evolutionary histories of the passive $z=0$ galaxy sample considered in section \ref{sec:sizeevo}.
However, restricting the analysis to galaxies that satisfy the compact galaxy criteria excludes many of the \highz\ passive galaxies due to its strong dependence on stellar mass (see dot-dashed curve in Fig. \ref{fig:compact_evo}). 
In the following analysis we consider the evolution of all \highz\ passive galaxies, not only those that are compact, but we emphasise the evolution of the compact sub-sample.

We select all galaxies that are identified as passive in at least one output in the redshift range $1.5 < z < 2$ (corresponding to the bottom right panel of Fig. \ref{fig:size_mass}).
We ensure no galaxy is double counted if identified as passive in multiple outputs in this redshift range by ensuring we include only the highest-redshift galaxy identified as passive on any tree branch.
We identify 331 \highz\ passive galaxies among the $\sim 10,000$ galaxies with $M_* > 10^9$\msun\ in this redshift range\footnote{The number of galaxies with $M_* > 10^9$\msun\ increases from 7980 at $z=2$ to 10510 at $z=1.5$ due to the formation and mass growth of galaxies.}.
Of the 331, 44 are and remain the central galaxy of their halo until $z=0$, 83 are or become satellite galaxies by $z=0$, while the remaining 204 merge with a more massive galaxy\footnote{If a galaxy is not on the main branch of a $z=0$ galaxy it is considered to have merged with a more massive system.}.
Focusing specifically on compact \highz\ passive galaxies, 26 of the 331 are compact at $1.5 < z <2$, 9 are centrals at $z=0$, 7 are satellites and the remaining 14 are merged into other systems. 
With over 60\% of all \highz\ passive galaxies and $\sim 50\%$ of compact galaxies merged with more massive systems by $z=0$, the size evolution of their descendant is not set primarily by the properties of the passive or compact galaxy, thus rendering these properties largely inconsequential to the $z=0$ population.

Fig. \ref{fig:compact_evo} shows the evolution of the \highz\ passive galaxies in the size-mass plane for the central and compact selections.
All galaxies from the central \highz\ passive population increase in size between $z=2$ and 0, with most galaxies also increasing in mass.
The resulting $z=0$ descendants have grown towards the present-day \smr.
Although these galaxies were selected to be passive in at least one output in the range $1.5<z<2$, 25 of the 44 are active at the present day (green symbols).

The growth of the compact galaxy selection is highlighted in Fig. \ref{fig:compact_evo} using diamond symbols and black lines. 
These galaxies follow the trends of the central population, typically growing in both mass and size, with some experiencing renewed star formation.
Note that none remain compact until $z=0$.

A histogram of the size growth for centrals, satellites and compact galaxies is shown in Fig. \ref{fig:gdist}.
The \highz\ passive galaxies that remain as centrals to $z=0$ increase in size with a median size growth of a factor of 4; the growth ranges from $\sim 2$ to 30.
Galaxies that are identified as satellites at $z=2$ undergo a variety of size growth scenarios; some of these galaxies increase in size by up to a factor of 10, while others decrease in size, with a median size growth of $\sim 1$.
A decrease in galaxy size can result from environmental processes, such as stripping and harassment \citep{Moore96}, experienced by galaxies in dense environments.
The compact galaxies, which include centrals, satellites and merged systems, all increase in size, by factors of $\sim 2$ to 40, although those experiencing size growth $> 20$ have been merged with a more massive system.

\subsection{Growth mechanisms of compact galaxies}
\label{sec:compact}

\begin{figure}
  \centering
  \includegraphics[width=0.5\textwidth]{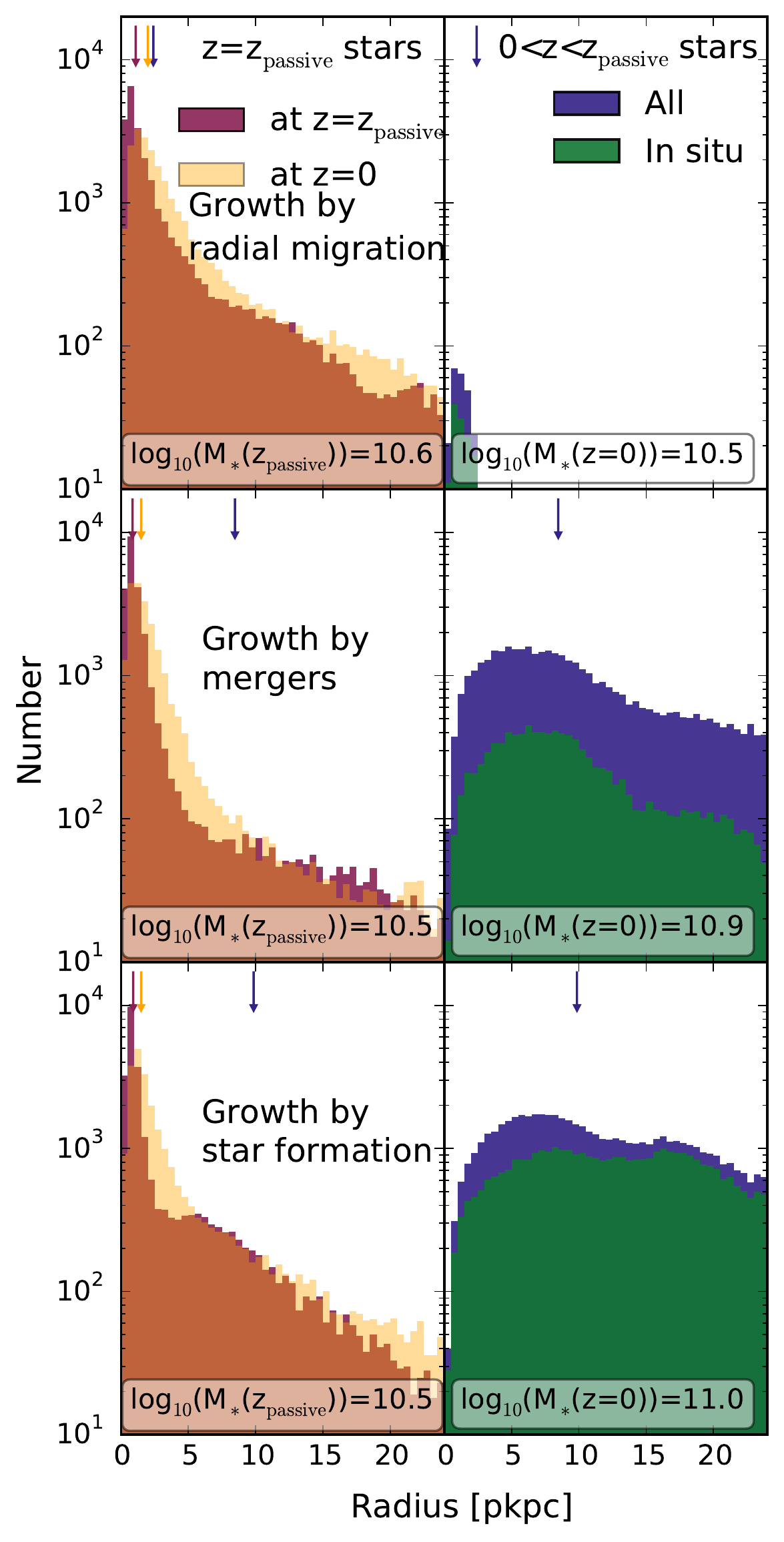}
  \caption{
The radial distributions of star particles in 3 example galaxies.
In the left panels, red histograms show the distribution of star particles in the galaxy at $z=2$, while the yellow histograms show the distributions of the same star particles at $z=0$.
In the right panels all other star particles in the $z=0$ galaxies are shown in blue.  
Those which formed in situ are shown in green.
The scales in the panels are the same, enabling a direct comparison of the numbers of star particles of each type.
The sizes of the galaxies at $z=2$ 0 are shown as red and blue arrows respectively.
The yellow arrows show the half-mass radii of the $z=2$ star particles at $z=0$.
In all cases the $z=2$ star particles have migrated to larger radii by $z=0$ resulting in galaxy size growth.
Further size growth is contributed by accreted stars in the middle panels and by in situ formation of stars in the lower panels.
	}
 \label{fig:rdist}
\end{figure}

\begin{figure}
  \centering
\minipage{0.23\textwidth}
  \includegraphics[width=1.0\textwidth]{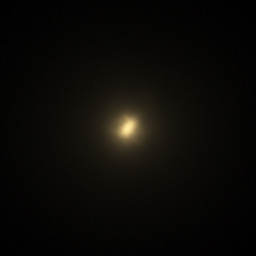}
\endminipage\hfill
\minipage{0.23\textwidth}
  \includegraphics[width=1.0\textwidth]{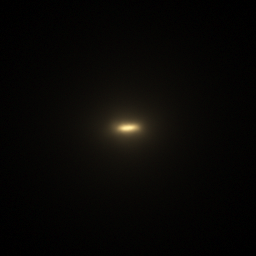}
\endminipage\hfill
\minipage{0.23\textwidth}
  \includegraphics[width=1.0\textwidth]{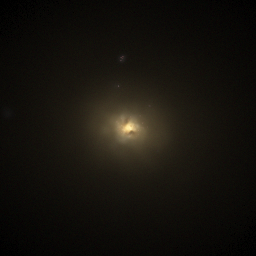}
\endminipage\hfill
\minipage{0.23\textwidth}
  \includegraphics[width=1.0\textwidth]{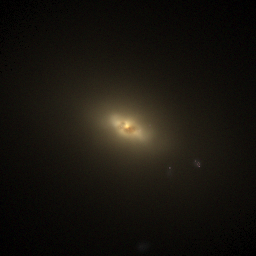}
\endminipage\hfill
\minipage{0.23\textwidth}
  \includegraphics[width=1.0\textwidth]{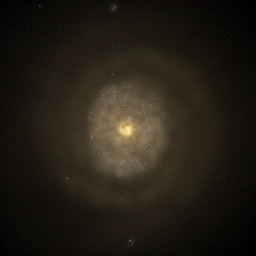}
\endminipage\hfill
\minipage{0.23\textwidth}
  \includegraphics[width=1.0\textwidth]{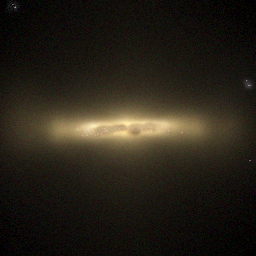}
\endminipage\hfill
  \caption{
Images of galaxies from Fig. \ref{fig:rdist} at $z=0$ in face-on and edge-on.
Top, middle and bottom rows correspond to the galaxies shown in the top, middle and bottom rows of Fig. \ref{fig:rdist} respectively.
The images were produced using the radiative transfer code \skirt\ \citep{Baes11} applied to \Eagle\ galaxies as described in \citet{Trayford15b}.
Stellar light in {\it g}, {\it r} and {\it i} band SDSS filters is shown, with dust attenuation.
Images are shown in thumbnails of 60 \kpc\ per side. 
The galaxy in the top row increases in size due to the redistribution of its $z=2$ stars, the middle row galaxy grows by a larger factor due to accreted stellar mass, while the bottom row galaxy experiences renewed star formation in the outskirts of the galaxy forming at disk at $z=0$.
	}
 \label{fig:images}
\end{figure}

We extend the analysis of the compact galaxies to consider how they grow.
In Fig. \ref{fig:rdist} we show the radial profile of 3 example galaxies, chosen to illustrate different growth mechanisms.
The left panels show the distribution of the star particles at $z=2$, and of these same particles at $z=0$. In all cases the stars migrate radially outwards. The red and yellow arrows compare the half-mass radii of these stars at $z=0$ and 2 respectively.
The half-mass radii of the galaxies at $z=0$ are indicated by a blue arrow. 
Comparing the yellow and blue arrows shows how much of the total size growth is due to the migration of $z=2$ stars relative to other sources of size growth such as accretion or in situ star formation.
The right panel shows the radial distributions of stars formed since $z=2$, with those formed in the main progenitor of the galaxy highlighted in green.
Fig. \ref{fig:images} shows images of these three galaxies at $z=0$ in edge-on and face-on projections.

The top panels show a galaxy whose size growth can almost entirely be accounted for by the radial migration of its $z=2$ star particles.
The \hmr\ of $z=2$ stars at $z=0$, \hmrz, is approximately equal to $R_{50}(z=0)$ (yellow and blue arrows respectively).
The star particles added to the system at $z<2$, shown in the right hand panel, contribute less than 10\% of the $z=0$ stellar mass and are centrally concentrated.
Similar galaxies, with \hmrz\ $\geq 0.5R_{50}(z=0)$ account for 13 of the 26 compact galaxies.  
In 6 of these galaxies the $z=2$ stars dominate the $z=0$ stellar mass, while in one further case the $z=2$ stars account for 40\% of the mass.
In the remaining 6 cases the compact galaxy is not the main progenitor of its $z=0$ descendant and there is significant evidence for mixing of the $z=2$ star particles in these cases.

The middle panels show a system where \hmrz\ $\ll R_{50}(z=0)$.
The origin of the additional growth in size of this system can be understood from the right-hand panel.
In this galaxy many star particles have been accreted (difference between blue and green histograms), particularly at large radii, driving the size growth.  
The accreted particles account for $\sim 40\%$ of $M_*(z=0)$, with stars formed in situ accounting for a further $\sim 20\%$ and $z=2$ star particles (after accounting for stellar mass loss) accounting for the remaining 40\%.
The accreted stars do not dominate the stellar mass, but their radial distribution significantly increases the \hmr.

Not all galaxies with \hmrz\ $\ll R_{50}(z=0)$ grow to their $z=0$ sizes due to accretion and mergers.
In some cases star formation is a major contributor, as seen in the bottom panels of Fig. \ref{fig:rdist}.
This star formation may be triggered by mergers, but the stars were formed in situ.
Note that we cannot tell from this plot whether the stars formed at large radii or whether they migrated outwards after their formation.
However images of this galaxy at $z=0$, as seen in the bottom panel of Fig. \ref{fig:images}, indicate that a disk of recently formed stars is in place in the outskirts of the galaxy.

It is evident from the simulation that the growth mechanisms of compact galaxies are diverse.  
All compact galaxies increase in size, but some only grow through radial migration, limiting their size growth to factors of $\sim 2$, while others grow through mergers and/or renewed star formation, where the compact galaxy progenitor can be found at the core of the $z=0$ descendant. 
However, over 50\% of the compact galaxies identified at $z \sim 2$ have merged into more massive systems by $z=0$ and their $z=0$ descendants therefore do not obviously reflect the compact galaxies' histories.
While it is tempting to conclude that the majority of compact galaxies merge into more massive systems by $z=0$, the shortfall in the number density of compact galaxies in the simulation precludes such a conclusion.
Likewise, the relative importance of the three growth mechanisms for compact galaxies in the simulation may not be accurate.

In summary, 60\% of the \highz\ passive galaxies merge into more massive systems by $z=0$, 25\% survive as satellites to $z=0$ and the remaining 15\% are centrals at $z=0$.
Most centrals and $\sim 50\%$ of the satellites that survive to $z=0$ increase in size with many of the centrals experiencing renewed star formation.
The compact sample of the \highz\ passive galaxies undergoes size growth up to a factor of $\sim 40$, growing through stellar migration (which, however, yields only modest size growth), renewed star formation and mergers.

%% file: Conculsions.tex
\section{Summary and Conclusions} 
\label{sec:summary}

We have studied the evolution of the galaxy \smr\ in the \Eagle\ cosmological hydrodynamical simulation, distinguishing between actively star forming and passive galaxy populations.  By comparing the half-mass radii from the simulation with the sizes inferred from the observations of \cite{vanderWel14}, we find (Figs. \ref{fig:size_mass}, \ref{fig:size_evo})
\begin{itemize}
\item The dependence of the sizes of simulated galaxies on stellar mass, redshift and star formation classification is close to that of observed galaxies. Galaxy sizes typically increase with stellar mass and decreasing redshift, and active galaxies are typically larger than their passive counterparts at a given stellar mass.
\item The level of agreement between the predicted and observed sizes is of order 0.1 -- 0.2 dex below $10^{11}$\msun\ (equivalent to 1-2 \kpc) for $0 \leq z \leq 2$.  
\item The scatter in the sizes at a given mass is also similar in simulated and observed galaxies.
\end{itemize}

Since the simulation agree relatively well with the observational data, we used the simulation as a tool to explore the size evolution of galaxy populations. 
Our conclusions are as follows:
\begin{itemize}
\item We find that the scatter in the \smr\ correlates well with the specific star formation rate for active galaxies and with the time since assembly for passive galaxies (Fig. \ref{fig:scatter_1}).
Active galaxies of a given mass with higher specific star formation rates are typically larger.
Similarly, passive galaxies at a given stellar mass that assembled more recently are larger than those that formed earlier.

\item We compared the evolution of star forming and passive galaxies as determined from the redshift dependence of the \smr\ of the ensemble population (which can be accessed directly by observations) and the typical size evolution of individual galaxies based on their main progenitors (which can be traced explicitly in simulations through galaxy merger trees). 
For galaxies that are active at $z=0$ the evolution of the \smr\ for the population is similar to the typical evolution of individual objects.
However, contrary to the rapid evolution of the \smr\ for the population of passive galaxies, the size evolution of individual present-day passive galaxies is comparable to that of their star forming counterparts (Figs. \ref{fig:size_evo}, \ref{fig:deltaR_m}). 
This difference arises simply because most present-day passive galaxies were star forming at higher redshifts.

\item The relative growth of galaxy sizes from $z=2$ to $z=0$ increases strongly with present-day stellar mass (Fig. \ref{fig:deltaR_m}). This dependence is partly due to the influence of galaxy mergers on sizes: galaxies with a higher fraction of accreted mass relative to the median at fixed mass experience an excess in size growth (Fig. \ref{fig:mergers}).

\item We investigated the evolution of 313 galaxies identified as passive in at least one output in the redshift range $1.5<z<2$. Of these galaxies, 65\% merge into more massive objects by $z=0$, 26\% are satellites and 14\% are centrals at $z=0$.
Satellite galaxies experience diverse size growth, with many decreasing in both size and mass most likely due to environmental processes such as tidal stripping.
The central galaxies all increase in size between $z=2$ and 0 and at $z=0$ are found to lie near the present-day \smr s for the general population.
In $\sim 50\%$ of central galaxy cases, they experience renewed star formation, indicating that instantaneous passive identification at \highz\ is not indicative of a galaxy's future evolution.

\item Of the \highz\ passive galaxies, 26 were identified as compact using the criterion of \cite{vanderWel14}.
Of these galaxies, 14 merge with more massive systems, 7 are satellites and 9 are centrals at $z=0$.
For all compact galaxies, the $z=2$ stars migrate to larger radii by $z=0$.  
In 13 cases further growth occurs due to mergers and star formation; the driver of the growth depends on the radial distribution of stars added to the galaxy at $z<2$.
\end{itemize}

We have demonstrated the success of the primary \Eagle\ simulation in producing realistic galaxy sizes across cosmic time, while investigating the passive \highz\ population further.
However, we caution that we cannot rule out that the sizes of the
smallest galaxies ($\lesssim 1$ \kpc) are boosted by the polytropic equation of
state that is imposed on the unresolved interstellar medium.
Higher-resolution simulations that include the physics required to model
the cold interstellar gas phase are required to make further progress.

%% file: Appendix.tex
\section{Comparison with other simulations}
\label{ap:comp}
\begin{figure}
  \centering
  \includegraphics[width=0.5\textwidth]{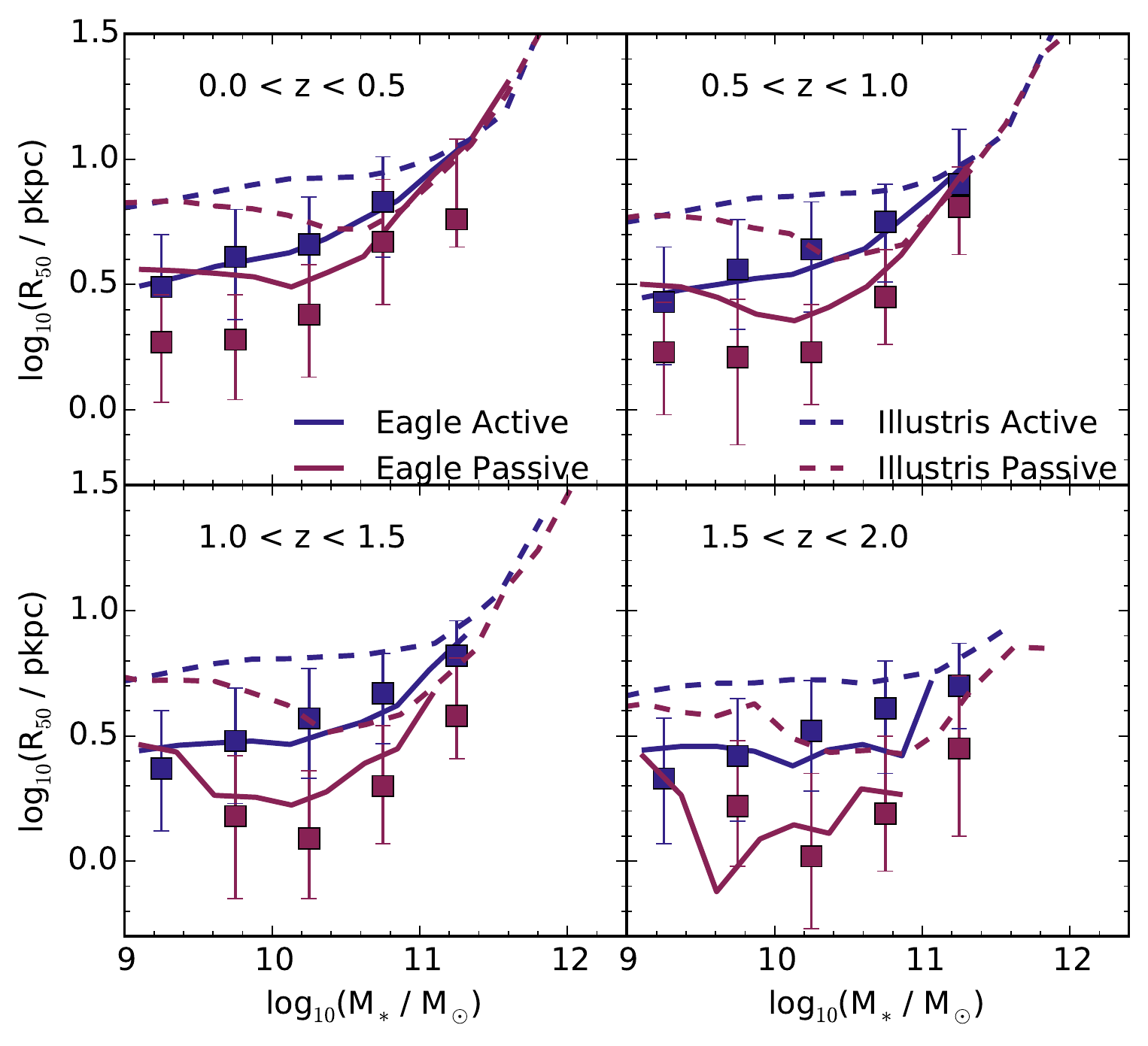}
  \caption{
The median \smr\ from \Eagle\ (solid curves) and \Illustrius\ (dashed curves) galaxy formation models.
Blue colours correspond to active galaxies, red curves to passive galaxies.
Observational data from \citet{vanderWel14} and \citet{Shen03} are shown as squares as in Fig. \ref{fig:size_mass}.
	}
 \label{fig:simcomp}
\end{figure}

In Fig. \ref{fig:simcomp} the \smr s from the \Eagle\ and \Illustrius\ galaxy formation simulations are compared in 4 redshift bins between 0 and 2, separating each galaxy population into active and passive galaxies using the SSFR cut described in section \ref{sec:ssfrcut}.
The \Eagle\ simulation is as in Fig. \ref{fig:size_mass}.
For \Illustrius\ \citep{Illustrius} the half-mass radii, stellar masses and star formation rates were obtained from their public data release \citep{Nelson15}.
An aperture of twice the half-mass radius is used measure to the stellar masses and star formation rates as was applied in previous \Illustrius\ publications.

For both hydrodynamical simulations, galaxy sizes tend to increase with stellar mass and there is an offset between the sizes of active and passive galaxies in all redshift bins shown.
Interestingly, the overlap in galaxy sizes of active and passive populations at $M_* < 10^{9.5}$\msun\ and $M_* > 10^{10.5}$\msun\ is seen in both simulations (as discussed in section \ref{sec:obs}).
While the trends are similar for both models, there is a normalisation offset in the sizes.
For $M_* < 10^{10}$\msun\ the \Illustrius\ galaxies are typically a factor of two (0.3 dex) larger than \Eagle\ galaxies and than observed.

\section{Resolution Tests}
\label{ap:res}

\begin{figure}
  \centering
  \includegraphics[width=0.5\textwidth]{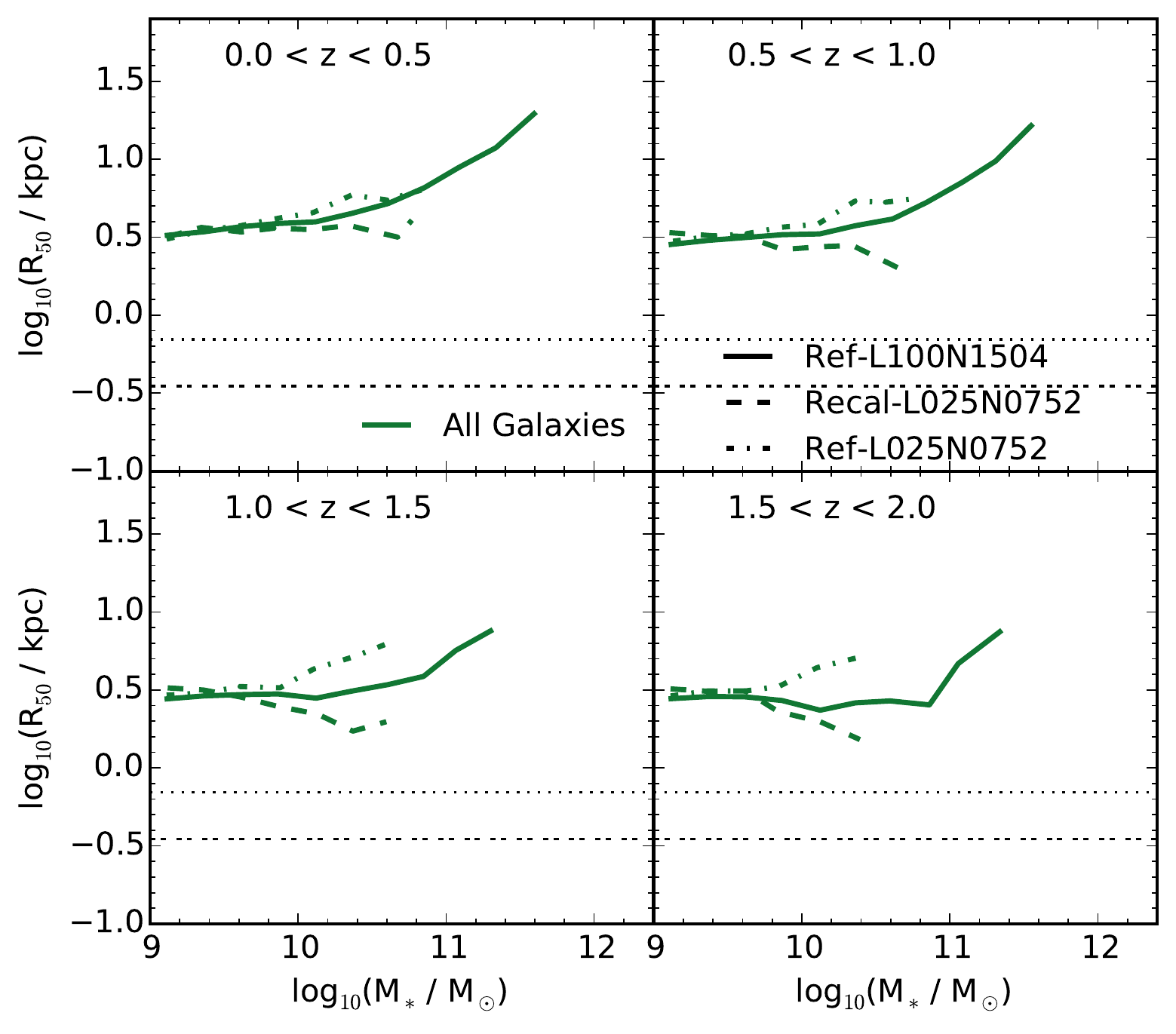}
  \caption{
Comparison of sizes for galaxies from \bigbox, \highresref\ and \highres\ over the range $0<z<2$ using the redshift bins on Fig. \ref{fig:size_mass}.
The solid, dashed and dot-dashed curves correspond to the median \smr s of \bigbox, \highres\ and \highresref, respectively.
The horizontal dotted and double-dot lines indicate the gravitational force softening of the intermediate and high resolution \Eagle\ simulations, respectively.
For the lowest redshift bin good convergence is seen between the models, increasing to higher redshift good convergence is maintained for $M_* < 10^{10}$\msun, but above this mass the sizes diverge.
	}
 \label{fig:res}
\end{figure}

In Fig. \ref{fig:res} strong and weak convergence tests for the evolution of the galaxy \smr\ are shown. 
Three simulations are presented: \bigbox, which was used in the main sections of the paper, \highresref, using the same subgrid physics parameters as \bigbox\ but run in a 25 cMpc box with 8 times better mass resolution and \highres, also in a 25 cMpc box with 8 times the mass resolution but with recalibrated stellar feedback and AGN parameters to better reproduce the local stellar mass function. 
The details of the recalibration of the high-resolution run can be found in \cite{Schaye15}. 

Comparing \bigbox\ with \highresref\ tests the strong convergence of the simulation (this is the typical convergence test carried out on hydrodynamical simulations), while comparing with \highres\ tests the weak convergence, a concept introduced by \cite{Schaye15} to account for the necessity of using weakly constrained subgrid physics prescriptions. 
Note that more massive galaxies form in the (100 cMpc)$^3$ simulation than the (25 cMpc)$^3$ simulations due to the larger box size.

In the lowest redshift bin, $0<z<0.5$, reasonable agreement is seen between all simulations, although the recalibrated model, \highres, exhibits a downturn at $M_* \gg 10^{10}$\msun.
At $z>1$ the \smr s at $M_* >10^{10}$\msun\ of the 25 cMpc simulations diverge, with the \bigbox\ simulation falling in between.

In the most massive galaxies in \highres\ the weaker density dependence of the subgrid model for feedback from star formation, which was introduced to better reproduce the galaxy stellar mass function, allows more compact galaxies to form at high redshift relative to \bigbox.
Meanwhile, in the \highresref\ simulation massive galaxies are larger because the feedback is more efficient in high-density regions.

\section{Variations in galaxy size definition}
\label{ap:sizes}

\begin{figure*}
  \centering
\minipage{1.0\textwidth}
  \includegraphics[width=1.0\textwidth]{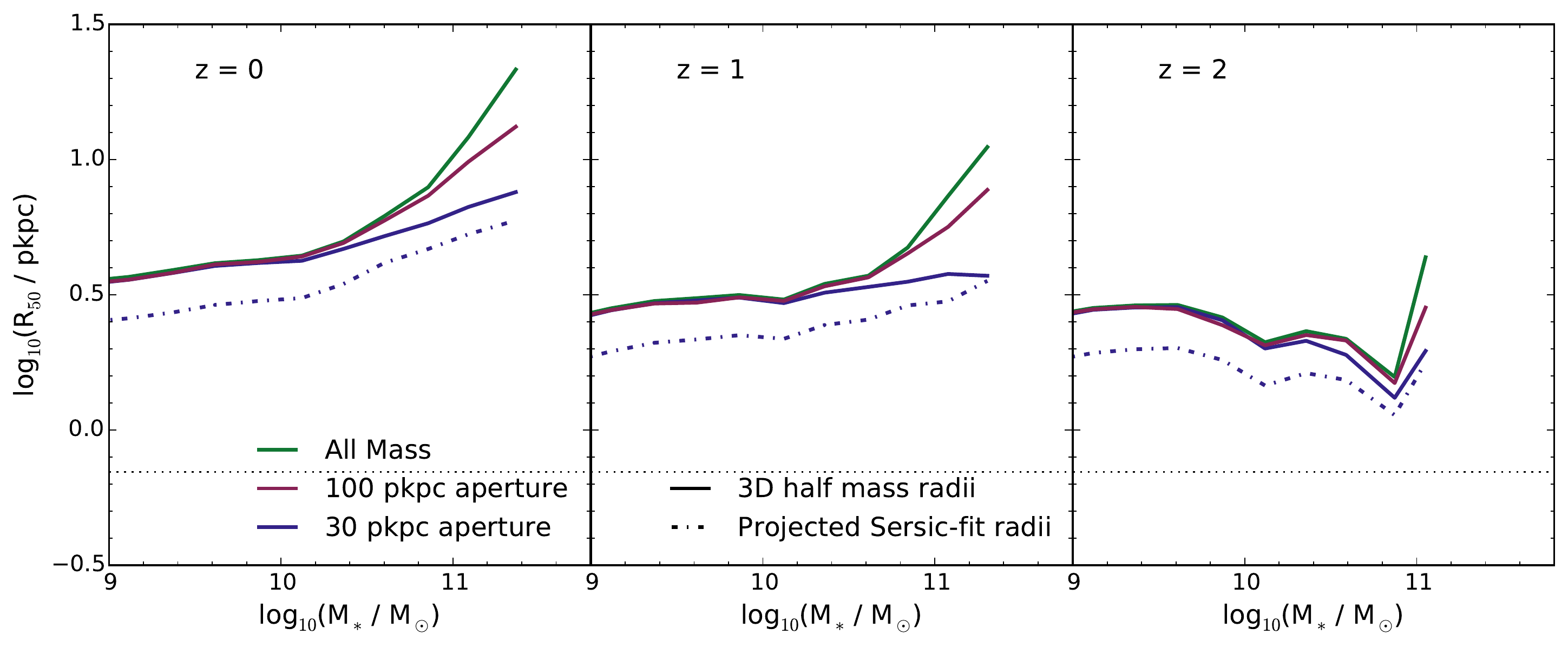}
\endminipage\hfill
  \caption{
The median \smr\ for galaxy sizes computed within different apertures and applying different measurement techniques at $z = 0, 1$ and 2.
Solid curves show the 3-D half-mass radius and dot-dashed curves show the projected half-mass radius based on a \sersic\ fit to the mass profile.
Green curves shows the radius based on all star particles bound to the subhalo, while red and blue curves only consider stars within 3-D 100~\kpc\ and 30~\kpc\ apertures respectively.
The solid red curves corresponds to the sizes used in this work, while the blue dot-dashed curves indicate the sizes used by \citet{Schaye15}.
The horizontal dotted line shows the gravitational force softening length in the simulation.
The aperture measurements affect the galaxy sizes at stellar masses $> 10^{10.5}$\msun, with galaxy size increasing with increasing aperture.
The different size measurement techniques produce similar trends with stellar mass, but there is a small systematic offset in size.
Note that we define the size as the 3-D half-mass radius in a 100~\kpc\ aperture.
    }
 \label{fig:sizedef}
\end{figure*}

We look at the variations in galaxy sizes due to the 3-D spherical aperture within which the total galaxy mass is measured and the effects of projection.
In Fig. \ref{fig:sizedef} the variation of the median galaxy \smr s recovered by applying different size definitions are compared at $z = 0, 1$ and 2.
The \hmr\ computed based on the total stellar mass and the mass within 3-D 100~\kpc\ and 30~\kpc\ apertures yield similar galaxy sizes for stellar masses below $10^{10.5}$\msun\ \citep[as is the case for the mass, see][]{Schaye15}, but larger apertures result in larger sizes above this mass.
In this paper the size is defined as the half-mass radius computed based on the mass within 100~\kpc\ apertures.
This aperture very closely replicates the half-mass radius determined by including all particles associated with a subhalo for galaxies with $M_* < 10^{11}$\msun.
However, it excludes stars far out in the most massive halos in the simulation, which would be difficult to observe (and are labelled intra-cluster light).

As well as comparing the effect of different apertures, we compute projected radii and compute the \hmr\ based on \sersic\ fits to the mass profile.
These variations result in systematic offsets in the \smr\ relative to the 3-D \hmr\ measurement.  
Due to the systematic nature of these offsets, we present only the 3-D \hmr\ measurement throughout, as variations in the definition do not impact the trends in galaxy sizes from the simulations.
We note, however, that the definition of galaxy may result in a systematic decrease of up to 0.2 dex.
A similar systematic offset has been found when computing observed radii using circularised and non-circularised methods \citep{Dutton11}.

\section{Definition of passive galaxies}
\label{ap:cuts}

In this appendix we discuss the impact of the definition of active and passive galaxies on the resulting difference in sizes reported between these two galaxy types (Fig. \ref{fig:size_mass}).
As discussed in section \ref{sec:ssfrcut}, the active / passive definition is based on the SSFR of a galaxy.
In observations, however, where star formation rates must be inferred, the definition of active and passive galaxies often depends on the observed colours of galaxies \citep[e.g.][]{vanderWel14}.
To determine if any significant difference arises by applying a colour cut, as opposed to a SSFR cut, at $z=0, 1$ and $2$ we produce UVJ colours for all simulated galaxies with $M_* > 10^9$\msun\ and compare the median \smr s.

Galaxy colours from the simulation are produced using the stellar population synthesis models of \cite{BC03} and applying the dust model of \cite{Charlot&Fall00}. 
A full description of the method can be found in \cite{Trayford15}. 
We apply two colour cuts to the UVJ plane, that of \vdw\ and of \cite{Muzzin13}.
We include the colour cut of Muzzin et al because at low redshift, where the \vdw\ sample is small (as commented on in section \ref{sec:obs}), the \vdw\ cut applied intersects the active or blue galaxy locus, resulting in an inappropriate division into active and passive galaxies\footnote{Note that \cite{Trayford15} showed that at $z=0.1$ the galaxy colours produced by \Eagle\ are in good agreement with observations, so for a larger observational sample a similar issue may arise when applying the UVJ cut of \vdw.}.

\begin{figure*}
  \centering
\minipage{1.0\textwidth}
  \includegraphics[width=1.0\textwidth]{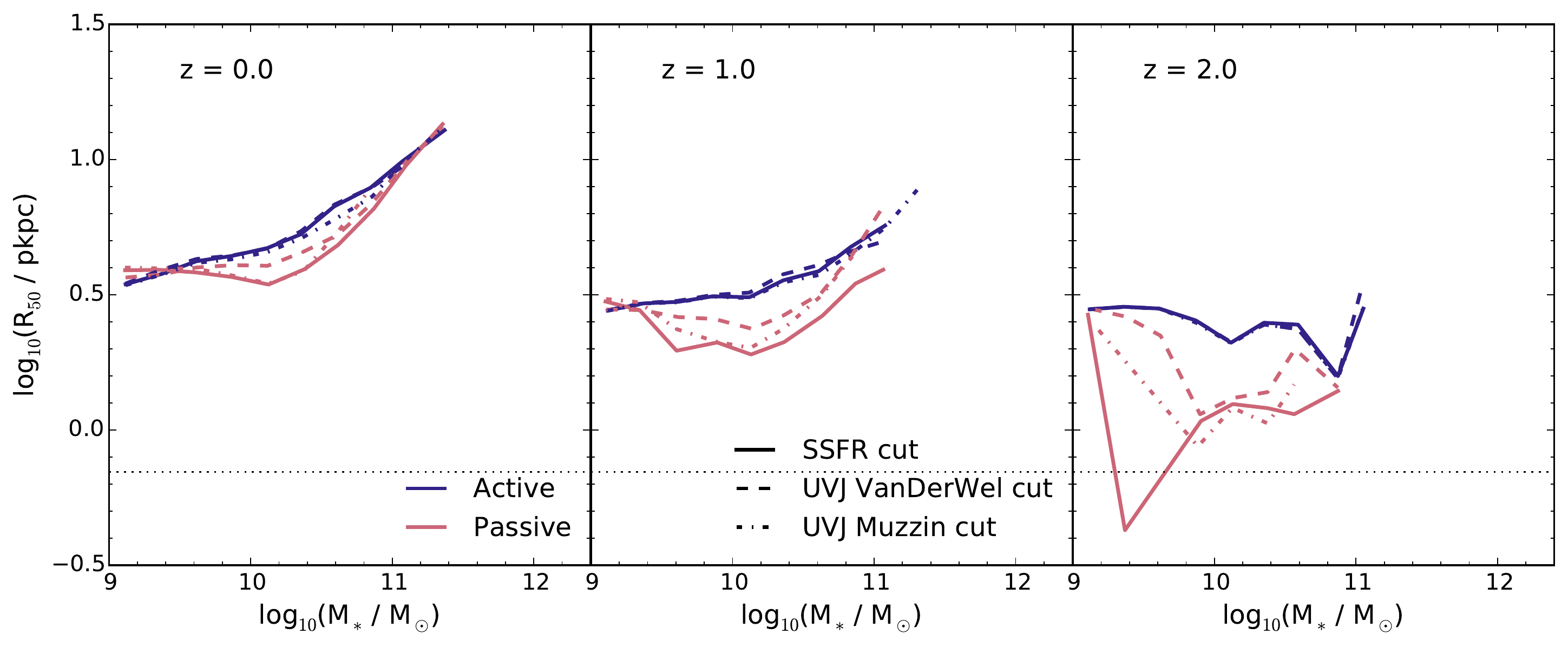}
\endminipage\hfill
  \caption{
The median \smr\ for different definitions of galaxy size at redshifts 0, 1 and 2 using 3 separations of active and passive galaxies.
The dotted line corresponds to the gravitational softening of the simulation.
At redshift 0 there is good correspondence between the different definitions.
At redshifts 1 and 2, differences appear for the passive galaxies, due to the different numbers of galaxies identified by each method, however, the offset in size between active and passive galaxies is evident for all three definitions.
Note that the noise in the $z=2$ passive galaxy sample is due to the small number of passive galaxies in the simulation at such a high redshift.
	}
 \label{fig:ssfr_cuts}
\end{figure*}

In Fig. \ref{fig:ssfr_cuts} the median \smr s for simulated galaxies at $z=0, 1$ and 2 are shown for all 3 definitions of active and passive galaxies: SSFR, UVJ applying the cut of \vdw\ and UVJ with the cut of Muzzin et al.
The median sizes of active galaxies are similar at all redshifts shown, irrespective of the details of the classification.
For passive galaxies there is always a clear offset from the sizes of active galaxies, but the median sizes of the passive galaxies show some variation between the cuts.
This is due to the different number of galaxies selected depending on the specifics of the cut.
Note that at $z=2$ the sample of passive galaxies selected from the simulation is small, as, for all cuts, the galaxies are predominantly active.
This causes some noise in the \smr\ at this redshift.

Note that the impact of the definition of passive galaxies on the passive fraction is explored in more detail in \cite{Trayford15b}.

In summary, how galaxies are classified results in negligible difference in the sizes of active galaxies, while the same is true for low-redshift passive galaxies.
For all cuts the active / passive galaxy size offset is maintained.

